\documentclass[nofootinbib,twocolumn,showpacs,preprintnumbers,amsmath,amssymb,aps]{revtex4}

\usepackage{graphicx} 
\usepackage{epsfig}
\usepackage{times}
\usepackage{amssymb}
\usepackage{amsmath}
\usepackage{color}
\usepackage{ulem}
\usepackage{textcomp}
\graphicspath{{.},{figs/}}%
\begin{document}
\title{Actin based propulsion: Intriguing interplay between material
properties and growth processes}

\author{Karin John}
\email{kjohn@spectro.ujf-grenoble.fr}
\affiliation{LSP UMR5588, Universit\'e J. Fourier, BP 87 - 38402 Grenoble
Cedex, France}
\author{Denis Caillerie}
\affiliation{L3S-R, BP 53 - 38041 Grenoble Cedex 9, France}
\author{Philippe Peyla}
\affiliation{LSP UMR5588, Universit\'e J. Fourier, BP
87 - 38402 Grenoble Cedex, France}
\author{Mourad Ismail}
\affiliation{LSP UMR5588, Universit\'e J. Fourier, BP 87 - 38402
Grenoble Cedex, France}

\author{Annie Raoult}
\affiliation{Laboratoire MAP5 UMR 8145, Universit\'e Paris
Descartes, 45 rue des Saints P\`eres, 75270 Paris Cedex 06, France}
\author{Jacques Prost}
\affiliation{Physico-Chimie, Institut Curie UMR168, 26 rue d'Ulm, 75248 Paris Cedex 05, France}
\author{Chaouqi Misbah}
\email{cmisbah@spectro.ujf-grenoble.fr}
\affiliation{LSP UMR5588, Universit\'e J. Fourier, BP 87 - 38402 Grenoble
Cedex, France}
\begin{abstract}
Eukaryotic cells and intracellular pathogens such as bacteria or
viruses utilize the actin polymerization machinery to
propel themselves forward.
Thereby, the onset of motion and choice of direction may be the result
of a spontaneous symmetry-breaking or might be triggered by external
signals and preexisting asymmetries, e.g.\ through a previous septation
in bacteria.

Although very complex, a key feature of cellular motility is the
ability of actin to form dense polymeric networks, whose
microstructure is tightly regulated by the cell.
These polar actin networks produce the forces necessary for propulsion
but may also be at the origin of a spontaneous symmetry-breaking.

Understanding the exact role of actin dynamics in cell motility
requires multiscale approaches which capture at the same time the
polymer network structure and dynamics on the scale of a few
nanometers and the macroscopic distribution of elastic stresses on the
scale of the whole cell.

In this chapter we review a selection of theories on how mechanical
material properties and growth processes interact to induce the onset
of actin based motion.
\end{abstract}
\pacs{
87.17.Jj 
87.15.Rn 
82.39.-k 
82.35.Pq 
62.40.+i 
81.40.Jj 
}
\maketitle

%
%
\section{Introduction}
%
%
Most living cells are able to perform a directed motion, either by
swimming in a liquid environment, by crawling on a solid support or by
squeezing through a three-dimensional matrix of fibers.

The speed of swimming bacteria can reach up to 100\,\textmu{}m\,s$^{-1}$,
whereas eukaryotic cell crawling can be as fast as 1\,\textmu{}m\,s$^{-1}$
(Ref.\,\cite{FlT04} and references therein). Given their size and
speed, the motion of single cells is governed by viscous forces, not
inertia, i.e.\ the Reynolds number $Re\ll 1$.

Unicellular organisms move in search of a food or light
source. Intracellular pathogens like bacteria or viruses spread by
exiting their host cell and entering a neighboring cell.
Other simple organisms like the slime mold {\it Dictyostelium
discoideum} migrate under unfavorable conditions, e.g.\ starvation,
towards an aggregation center to form a multicellular organism.

Typically, in a multicellular organism not all cells are motile all
the time but they can be mobilized by the appropriate stimuli. For
example, the ability to move plays a crucial role during embryonic
development, in wound healing and in the immune
response. Additionally, cellular motility is a prerequisite for
metastasis formation during cancer development

The biological realizations to produce a propulsive force are
diverse. Most swimming cells, e.g.\ sperm cells or the bacterium
{\it Escherichia coli}, use one or multiple beating flagella\footnote{long
cellular extensions}, respectively. In contrast, crawling cells and
some intracellular pathogens advance by actin polymerization.

In this article we will present recent experiments and concepts to
understand the latter mechanism, the production of forces in the
advancing edge of crawling cells or for the propulsion of
intracellular organelles, which is also of fundamental interest for
the medical and engineering sciences.\\
The foundation of the research of cell motility as a distinct
discipline were laid in the 1970's by the group around Michael
Abercrombie \cite{Abe80}.
He was the first to divide the motion of fibroblasts\footnote{most
common cells of connective tissue in animals} into three phases,
Extension - Adhesion - Contraction, which form the dogma of cellular
motion as it is recognized today.
In this mechanism, a slow movement ($\sim$1\,\textmu{}m\,min$^{-1}$) is
generated by the extension of flat membrane sheets, lamellipodia, into
the direction of movement. The advancement of the membrane is
accompanied by the formation of focal adhesions, contacts between the
substratum, the cell membrane and actin stress fibers. Finally the
cell rear retracts, accompanied by a deadhesion of the membrane from
the substratum.

However, there exist variations of this dogma. Fish
keratocytes\footnote{fish scale cells} perform a rapid continuous
motion ($\sim$10\,\textmu{}m\,min$^{-1}$) with a constant shape. They
almost seem to glide over the surface and form only transient focal
contacts with the substratum with a much shorter lifetime than the
focal adhesions formed in fibroblasts \cite{AnC00}.

Another variant is the rapid motion ($\sim$10\,\textmu{}m\,min$^{-1}$) of
the slime mold {\it Dictyostelium discoideum}, which moves in an
amoeboid fashion. During this amoeboid motion only non-specific
contacts with the substratum are formed and actin stress fibers are
absent \cite{FuI97}.

Despite the various mechanisms, cell motion requires first of all the
self-organization of the cell into an advancing and receding edge.
This manifests itself by different molecular concentrations or
activation levels of enzymes at the two poles.
The polarization can be guided by external signals, e.g.\ chemical
gradients or a variation in the mechanical properties of the support,
but it might also arise in a homogeneous environment, after a
transient mechanical perturbation of a stationary symmetric cell
\cite{VSB99}.

The three processes, extension, adhesion and contraction are coupled
by sophisticated and complex mechanisms. However, it seems as if the
process of front extension relies on a completely different machinery
than that of adhesion and contraction and can be studied separately.

\begin{figure}
\begin{center}
\includegraphics[width=0.7\hsize]{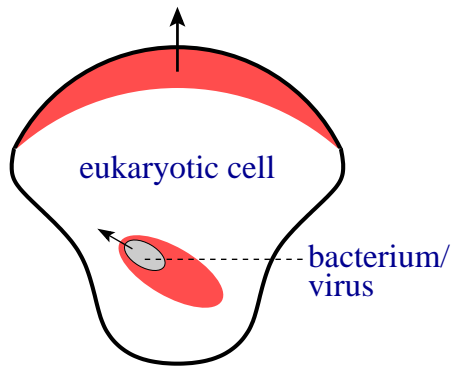}
\end{center}
\caption{Actin polymerizes into an elastic filament network (shaded in
grey) at the leading edge of motile eukaryotic cells or at the outer
surface of pathogens and organelles and induces cellular motion. The
direction of motion is indicated by the black arrows.\label{mot}}
\end{figure}

It had been known for a long time that on a cellular level, forces
can be generated on the basis of muscle like proteins, i.e.\ actin
and myosin, which is indeed responsible for the contraction of the
cell rear \cite{IRH76}.
However, more recently it was discovered that the protrusions at the
leading edge as well as the motion of intracellular organelles
\cite{KTD06} or pathogens, like the bacterium {\it Listeria monocytogenes}
\cite{TMT92} or the {\it Vaccinia} virus \cite{CCG95}, can be associated
with the so-called actin polymerization machinery \cite{ThM91} as
shown schematically in Fig.\ \ref{mot}.
The molecular basis and
minimal ingredients of actin polymerization forces are now very well
understood thanks to biomimetic experiments in the group of
Marie-France Carlier \cite{LBP99}.
However, the theoretical understanding of the physical mechanisms of
polymerization process are still a matter of debate and shall be the
main topic of this review.

In the first introductory part we will outline the biochemical basis
of actin polymerization and present a selection of experimentally
observed phenomena. Here we will shortly present the processes taking
place at the leading edge of locomoting cells and then we will mainly
rely on biomimetic experiments. In a second part we will present some
theoretical concepts to interpret and understand the existing
experiments. The list of presented models is by no means complete and
rather represents a selection of the leading ideas.
Here we will focus on the class of Brownian ratchet models and
macroscopic models on symmetry-breaking.
%
%
%
%
\section{Experimental observations}
%
%
%
\subsection{Biochemistry of actin polymerization and organization of the leading edge of advancing cells}
Actin is a small globular protein of 42\,kDa present in all eukaryotic
cells \cite{Stryer}. Under physiological conditions actin monomers
(G-actin) polymerize into long helical filaments (F-actin).
In a living organism these polymerization or depolymerization
processes are tightly regulated.
The literature on the dynamics of actin and the
proteins which interact with actin is vast.
As an introduction we would like to refer the reader to the
comprehensive biochemical reviews by Pollard
et. al. \cite{PBM00,PoB03} and Rafelski et al. \cite{RaT04} or the
books by Bray \cite{Bray} and Howard \cite{Howard}. Due to the limited
scope of this bookchapter we will just present the basic phenomena and
common terminology associated with the actin polymerization machinery,
which will allow the reader to understand the pertinent questions and
concepts.
\subsubsection{Actin polymerization in vitro}
%
%
Under physiological conditions, i.e.\ at an ionic strength of
$\sim$100\,mM, monomeric actin polymerizes spontaneously into
filaments.
The filament growth process typically starts with a nucleation
process, since actin dimers and trimers are unstable.
Shortening or elongation of existing filaments
occurs predominately via subunit addition or subtraction at the
filament ends and not via filament breaking or annealing processes.
Actin monomers at a concentration $c$ may bind to a filament end with
a rate $\sim k_+ c$ and dissociate with a rate $\sim k_-$. In a
stationary situation, i.e.\ zero net growth of the filament, one can
write
\begin{equation}
0=k_+c-k_-\,.
\end{equation}
The concentration $c_c=k_-/k_+$ associated with this equilibrium is
called critical concentration.

\begin{figure}
\begin{center}
\includegraphics[width=0.7\hsize]{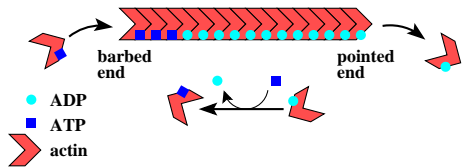}
\end{center}
\caption{Actin subunits treadmill through a filament. ATP-actin
polymerizes at the barbed end, ATP hydrolyzes and ADP-actin
depolymerizes at the pointed end. In the solution ATP is exchanged for
ADP at the monomers.\label{treadmill}}
\end{figure}

G-actin has a structural polarity \cite{HPG90}. The same polarity is
also found in actin filaments and can be visualized as an arrowhead
pattern by decoration of filaments with myosin\footnote{Myosin
molecules bind to each actin subunit in the filament in an oriented
fashion, leading to a typical pattern along the filament, which looks
like a series of arrowheads in an electron micrograph. For details see
Ref.\ \cite{Bray}.}. From this arrowhead pattern the two filament ends
are referred to as pointed and barbed, respectively, as shown in Fig.\
\ref{treadmill}.

In the presence of Mg-ATP the structural difference translates into
a difference of the critical concentrations and rate constants
between the barbed and pointed ends and causes a treadmilling
 of subunits through the filament.

Briefly, this phenomenon can be explained as follows (for details see
Ref.\,\cite{PBM00}). Most actin monomers are bound to ATP (typically
Mg-ATP). The critical concentration for this ATP bound species is
about 6 times lower for the barbed end than for the pointed end.
For ADP actin the critical concentrations are about the same for both
ends but about ten times higher than for ATP actin at the barbed end.
Therefore, in the steady state the ATP actin concentration is above
the critical concentration of the barbed end and below the critical
concentration of the pointed end.
Polymerized actin subunits are still bound to ATP but in the
course of time ATP hydrolyzes irreversibly into ADP+P$_i$ and, later
on, the anorganic phosphate P$_i$ dissociates from the filament with a
half-time of several minutes.
Consequently, ATP actin polymerizes at the barbed end, travels along
the filament whereby ATP is hydrolyzed and finally ADP actin
depolymerizes at the pointed end.
Fig.\ \ref{treadmill} shows schematically the main key processes of
the treadmilling cycle.

The process of irreversible ATP hydrolysis and P$_i$ release (and
subsequently the exchange ADP$\rightarrow$ ATP at the actin monomer in
solution) keeps the system out of equilibrium and allows for a
constant flux of monomers through the filament at constant filament
length, which is called treadmilling and forms the base of cellular
motility.
\subsubsection{Proteins that regulate actin polymerization in vivo}
%
%
In the living cell the actin polymerization machinery is tightly
regulated by signalling processes.
Many different factors interact with actin or participate in the
regulation of this polymerization machinery.
In the following paragraphs we will only discuss the relevant factors
which determine the actin architecture of the leading edge of
advancing cells or the actin comets formed by intracellular pathogens
and which seem to be crucial to generate motion.

First of all, most actin monomers are bound to so-called monomer
binding proteins, e.g.\ Thymosin-$\beta$4 and Profilin, and are thus
not able to nucleate new filaments. But Profilin-ATP-actin complexes
elongate existing filaments at the barbed end nearly as efficiently as
ATP-actin.

Electron micrographs of the leading edge show a dense network of actin
filaments linked to each other by Y junctions, where a new filament
grows from an existing filament at a 70$^\circ$ angle. The distance
between two crosslinks is of the order $\sim$20--30\,nm (in
comparison, the persistence length of actin filaments is 15\,\textmu{}m
\cite{OMS93}).
This Y junction is initiated by the interaction of the Wasp/Scar
protein and the Arp2/3 complex, whereby Wasp/Scar is a membrane bound
protein which incites the binding of the Arp2/3 complex to an existing
filament, which in turn then serves as a nucleation point for a new
filament.

Free barbed filament ends are quickly covered by so called capping
proteins, thus limiting filament elongation to a zone near the plasma
membrane.  At the pointed filament end depolymerization takes
place. This process can be accelerated by a protein complex called
ADF/cofilin, which is able to sever old filaments containing ADP actin
subunits. These filament fragments are then depolymerizing rapidly.
Besides Arp2/3 there exist other types of filament nucleators, called
formins. In the presence of profilin and ADF/cofilin they seem to
favor the growth and bundling of several actin filaments into cables
\cite{MBG07}, present in spike-like membrane extensions, called
filopodia. In this chapter we will limit ourselves to actin networks
produced by the Arp2/3 complex.

The above outlined process of polymerization at the membrane/actin
network interface and depolymerization far away from the membrane
provides the mechanism for pushing the membrane into the direction of
movement.
However, little is known about the mechanical properties of such
filament networks and recent experiments indicate that the loading
history determines the growth velocity of the network \cite{PCT05}.

Certain bacteria invade other living cells and hijack the actin
polymerization machinery of their host cells to propel themselves
forward. They carry a protein on their outer surface (e.g.\ ActA for
{\it Listeria monocytogenes}), which adopts the same function as the
Wasp/Scar protein in the membrane of eukaryotes: it triggers the
polymerization of an Arp2/3 crosslinked network at the outer bacterial
surface. This actin network typically develops asymmetrically only at
one side of the bacterium thus pushing the bacterium in the other
direction.

A similar mechanism might also be responsible for the motion of
endocytic vesicles in living cells (see \cite{KTD06} and references
therein).

To summarize the growth processes and the resulting architecture of
the actin system at the leading edge:
close to the cell membrane containing an activating enzyme (e.g.\ Wasp
or ActA), actin polymerizes into a dense crosslinked filament network,
which extends several \textmu{}m into the cell and where fast polymerizing
barbed ends are oriented towards the membrane. Polymerization is
restricted to a narrow zone close to the membrane since free barbed
ends are rapidly blocked by capping proteins.  Free pointed ends which
are far away from the membrane are depolymerizing.
These out of equilibrium growth processes driven by the irreversible
hydrolysis of ATP lead to the extension of membrane protrusions or
propel bacteria forward.
\subsection{Biomimetic experiments}
\subsubsection{General observations}
The minimal set of biochemical ingredients to induce actin driven
motion have been identified about ten years ago \cite{LBP99}. At the
same time actin driven motion has been successfully reconstituted in
vitro by replacing the bacterium by mimetic objects, e.g.\ beads
\cite{CFO99,NGF00,YTA99}, vesicles \cite{GFT03,MCJ98,UCA03} or
droplets \cite{BCJ04}.

%
%
In these experiments objects (hard, soft, fluid) coated with either
ActA or Wasp/Scar proteins are added to a solution containing ATP
actin, capping protein and a few well defined regulatory proteins.
With this design, actin polymerizes predominantly at the surface of
the object, i.e.\ the internal interface, and depolymerizes at the
interface between the network and the solution, i.e.\ the external
interface.

\begin{figure}
\begin{center}
\includegraphics[width=0.7\hsize]{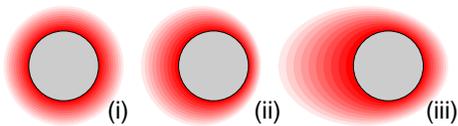}
\end{center}
\caption{Schematic view of the evolution of an actin gel (shades of
red) around a bead (grey) as described in the text with (i) a
symmetric growth, (ii) the symmetry-breaking, and finally (iii) the
comet formation.\label{symbreak}}
\end{figure}

After an initial phase (i), where polymerization occurs symmetrically
around the object, the symmetry is broken and the actin cloud starts
to grow asymmetrically (ii). In later stages an actin comet develops
(iii), and the object starts to move with velocities up to
0.1\,\textmu{}m\,s$^{-1}$.
A schematic representation of the three phases of the actin cloud
evolution is shown in Fig.\ \ref{symbreak}.
Interestingly the mode of movement depends on the surface parameters
of the object. Depending on the conditions one can observe a
continuous motion or a saltatory motion, where the object undergoes
stop-and-go cycles \cite{BWG02,DSR08,LGG97,TCS07}, which is also
reflected by variations in the actin density in the comet.

%
%
%
The network grown around these biomimetic objects has elastic
properties with a Young's modulus of 10$^3$-10$^4$\,Pa
\cite{GLO00,MPC04} and is often referred to as an actin gel.

The actin filaments interact, at least transiently, with the
activating enzyme bound to or adsorbed on the objects surface, e.g.\
the stress to detach an actin comet from a bead has been estimated
to be of the order of $\sim 100$\,pN\,\textmu{}m$^{-2}$ \cite{MPC04},
whereby the adsorbed activating enzyme stays on the
bead. Interestingly, in the same experimental setup actin comets under
compression appeared to be hollow.

In regime of continuous motion the bead velocity is not affected by
the viscosity of the medium (over five orders of magnitude). This
raises the question of the dissipative force, which is obviously not
the Stokes force on the object (about 1\,fN for a bead of
1\,\textmu{}m radius moving with a velocity of 1\,\textmu{}m\,min$^{-1}$ in a
standard motility assay). In contrast, the stalling force for the
growing actin tail is a few nN \cite{MPC04,PJL08}).
Several studies indicate that friction between the actin gel and the
propelled object is the major source of dissipation
\cite{GCR00,WHD03}. This would support the hypothesis, that the
saltatory bead or vesicle motion is the result of a stick-slip motion,
where a certain critical force has to be overcome to rupture bead-gel
bonds and to displace the object with respect to the gel \cite{BPS05,GCR00}.
\subsubsection{Symmetry-breaking}
Biomimetic experiments are not only an effective tool to study the
generation of motion by polymerization but they have also revealed a
spontaneous symmetry-breaking instability in the growing actin network
(transition form (i) to (ii) in Fig.\ \ref{symbreak}), whose
consequences in vivo are not clear and whose nature is still a subject
of debate in the literature.

Even though, as often argued by biologists, a symmetry-breaking does
not play a role in bacterial systems, since the ActA proteins are
distributed asymmetrically around the bacterium due to a previous cell
division, the instability is a powerful tool to study the coupling
mechanisms between actin polymerization and mechanical stresses.

So far there seems to be a consensus in the literature, that
mechanical stresses build up in the actin network due to growth, since
new monomers are inserted at the internal curved interface and push
older network layers away from the object. In a symmetric situation
one expects therefore high tangential stresses at the external
actin/solution interface and high normal stresses at the internal
object/actin interface.
Then the symmetry-breaking is driven by a release of elastic stresses
in the actin gel, either by an asymmetric
polymerization/depolymerization or by a fracture at the external gel
interface, whereby the two mechanisms are difficult to distinguish.

There seem to exist subtle differences in the nature of the
symmetry-breaking for hard \cite{CFO99,NGF00,YTA99} and soft objects
\cite{GFT03,MCJ98,UCA03}, as was demonstrated in a more recent study
\cite{DSR08}. In fluorescence labeling experiments it was shown that
for soft objects (vesicles) the symmetry is broken at the internal gel
interface, i.e.\ polymerization is considerably slowed down on one
side of the vesicle, such that an actin tail develops at the opposite
side.
In contrast, for hard beads the symmetry is broken at the external
(depolymerizing) gel interface, whereby some authors suggest, that the
filament network is ruptured due to an accumulation of tangential
stresses \cite{GPP05}, whereas theoretical models indicate that a
local stress dependent depolymerization is sufficient to induce a
symmetry-breaking \cite{JPK08,SPJ04}.
Note however, that actin driven motion itself does not require the
polymerization at curved surfaces, as has been demonstrated
theoretically \cite{Car01}.

Furthermore, the mechanical properties of the biomimetic object has an
effect on the mode selection of the instability. Whereas for soft
objects only instabilities, which produce one single actin tail, have
been reported, there are observations of higher order instabilities
for hard spheres depending on the experimental conditions
\cite{DSR08}. Therefore, the boundary conditions at the internal gel
interface seem to be crucial for the mode selection.
\section{Theoretical Approaches}
Two types of theoretical concepts, how polymerization processes
transform biochemical into mechanical energy, have been developing in
parallel over the past 15 years.

The first type of concept, called
``Brownian ratchet models'' was introduced by Peskin et
al. \cite{POO93}. Its major postulate is that polymerization
processes (e.g.\ actin polymerization) are able to rectify the
Brownian motion and can thus induce motion. Brownian ratchet models
describe microscopically the polymerization of actin
filaments in the presence of an obstacle.
Although very consuming in terms of computational efforts, Brownian
ratchet models allow for the incorporation of a very detailed
description of the kinetics of the polymerization machinery. They
provide ingenious tools to study the complex phenomena which
have been observed in biomimetic experiments.

The second type of concept, we shall refer to it as macroscopic
concept, does not focus so much on the dynamics of the single
filament, but rather considers the actin filament network as a
continuous elastic body under growth, where the growth dynamics is
driven by a thermodynamic force, the chemical potential. These
coarse-grained models emphasize the global stress distribution in the
filament network and the nonlocal aspect of elasticity, while neglecting the
details of the polymerization process.
Nevertheless, they are more suitable to describe and understand the
symmetry-breaking in actin gels around spherical objects in terms of
simple physical ingredients.

The ideal multiscale model would combine both concepts and use the
global stress distribution as an input for the complex polymerization
kinetics at the free interfaces.

In the following section we will briefly review the two concepts. In
the first subsection on Brownian ratchet we will mainly discuss Refs.\
\cite{EGF08,GFF08} as they provide, from our point of view, the most
advanced description for the polymerization dynamics close to the
obstacle. In the second subsection on macroscopic models we will
shortly outline and discuss the major drawbacks of Refs.\ \cite{SPJ04} and
\cite{JPK08} and then give a perspective, how these problems can be
solved using homogenization techniques.
\subsection{Brownian ratchet models}
\subsubsection{General concept}
As mentioned above, the idea of a Brownian ratchet model for
polymerization forces in biological system was introduced by Peskin
and coworkers \cite{POO93}.
This work explored the rectification of the Brownian motion of a
particle by the intercalation of new monomers at the interspace
between a filament tip and the obstacle, which gives an ideal ratchet
velocity $v$ of
\begin{equation}
v=\frac{2D}{\delta}\,,
\end{equation}
where $D$ denotes the diffusion coefficient of the obstacle and
$\delta$ the size of a monomer.
However, later on it was shown, that the actin driven motion was
relatively independent of the obstacle size \cite{GoT95} and
independent of the viscosity of the medium over several orders of
magnitude \cite{WHD03}, i.e. independent of $D$.
Therefore, the concept of Brownian ratchets was generalized to
``elastic Brownian ratchets'', where the tips of polymerizing
filaments undergo fluctuations, which induce a propulsive force on the
obstacle \cite{MoO96,OuT99} and to ``tethered ratchets'' \cite{MoO03},
which involves also the transient attachment of fluctuating filament
tips to the obstacle.
Typically in this description the polymerization rate constant $k_p$
is weighted by the load $f$ using Kramers theory
\begin{equation}
k_p=k^{max}_p e^{-f \delta/k_BT}\,,\label{loaddep}
\end{equation}
where $k^{max}_p$ is the rate constant at zero load, $\delta$
represents the gap size to intercalate a monomer, $k_B$ and $T$ denote
the Boltzmann constant and the absolute temperature, respectively.

The general framework has been used to quantitatively model the steady
motion of flat objects \cite {Car01}, lamellipodia and bacterial
motion \cite{MoO96,MoO03}. However, stochastic effects in the number
of polarizing filaments were necessary to break the symmetry in a
spherical gel around a bead homogeneously covered by ActA or Wasp,
whereas the global elastic stress distribution due to filament
crosslinking was neglected \cite{MoO03,OuT99} .

Other attempts to model biomimetic motility quantitatively by
explicitly modelling the filament dynamics in the actin tail
succeeded in obtaining the crossover from a continuous to a hopping
motion \cite{AlO04,BuM07}. However, the velocity oscillations were on
a time scale of ms with stepsizes of a few nm, as opposed to the
experimental oscillations on the scale of several min with stepsizes
on the \textmu{}m scale \cite{BWG02,DSR08,LGG97,TCS07}.

More recently, the tethered ratchet model of Mogilner et
al. \cite{MoO03} was submitted to a more rigorous treatment concerning
the polymer physics of the filament brush close to the obstacle
\cite{EGF08,GFF08}. Further away from the obstacle the cross-linked
gel is advancing with a so-called grafting speed (force dependent and
coupled to the brush length). In the following we will highlight the
basic features of this model, since it quantitatively reproduced
velocity oscillations.

\subsubsection{A quantitative model for velocity oscillation in actin-based motility}
As mentioned previously, the motion of a mutant form of the bacterium
{\it Listeria} (with a mutation in the ActA protein) is oscillatory and
shows remarkable temporal patterns \cite{GCR00,LGG97}: the bacteria
move very slowly during 30 to 100\,s, jump forward during a few
seconds and then slow down again abruptly. Such a periodic behavior,
consisting of long intervals of a slowly changing dynamics that
alternate with short periods of very fast transition, are found in
several chemical and biological systems and are known as relaxation
oscillations \cite{KrS01}.

Gholami et al. \cite{GFF08} and Enculescu et al. \cite{EGF08}
have developed a microscopic model based on the concept of tethered
elastic ratchets for actin-based motility. Their model consists of a
brush of growing actin filaments close to an object (the bacterium)
and describes the generation of forces and consequently the propulsion
of the object.
\begin{figure}
\begin{center}
\includegraphics[width=0.7\hsize]{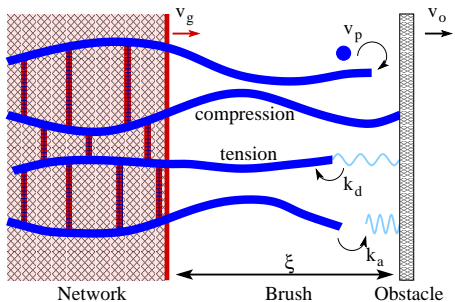}
\end{center}
\caption{\label{fig:scheme}Schematic representation of the elastic
ratchet model as explained in the text. Filaments are anchored with
one end into a cross-linked network forming an actin-gel and oriented
with the other end against an obstacle. Adapted from Refs.\
\cite{EGF08,GFF08}.}
\end{figure}

Briefly, the model considers the case of fluctuating filaments close
to an obstacle. Filaments may attach to the obstacle with a rate
constant $k_a$ and detach from the obstacle with a force dependent
rate constant $k_d$, resulting in two distributions of populations of
filaments, i.e.\ of attached and detached $n_a$ and $n_d$,
respectively.
Opposite to the obstacle the filament ends are anchored in a
cross-linked network, the actin gel.
Detached filaments polymerize with a load dependent velocity $v_{p}$.
The distance between the grafting point, i.e.\ the interface between
the network and the polymer brush, and the obstacle is denoted by
$\xi$. The filaments are characterized by their free contour lengths
$l$.
One of the crucial ingredients of the model is, that the grafting
point is advancing in the direction of the obstacle with a so called
grafting velocity $v_g$, which depends on the free contour lengths of
the polymers by
\begin{equation}
v_g(l)=v_g^{max}\tanh{(l/\bar l)}\,,
\end{equation}
where $\bar l$ denotes a characteristic width of boundary between the
crosslinked network and the filament brush.

Attached and detached filaments exert entropic forces \cite{GWF06} on
the obstacle, $F_{a}(l,\xi)$ and $F_{d}(l,\xi)$, respectively,
which lead to the propulsion of the object with an effective
friction coefficient $\zeta$.

The load dependence of the kinetic rate constants is again included
using a Kramers type expression [see Eq.\ (\ref{loaddep})], i.e.\
\begin{equation}
k_{d}(F_{a})  =  k_{d}^{0}\exp(-\delta\, F_{a}/k_{B}T)\,\label{eq:detachment rate}
\end{equation}
for the detachment and
\begin{equation}
v_{p}(l)=v_{p}^{max}\exp(-\delta F_{d}/k_{B}T)\,\label{eq:vitesse p}
\end{equation}
for the polymerization speed.

The full evolution of the length distributions of the two filament
populations $n_a(l,t)$ and $n_d(l,t)$ is described by
advection-reaction equations. However, it is shown that the two
distributions contract rapidly on the scale of 10$^{-2}$\,s into
monodisperse distributions $N_a=n_a(t)\delta(l-l_a(t))$ and
$N_d=n_d(t)\delta(l-l_d(t))$ localized at $l_a$ and $l_d$ for attached
and detached filaments, respectively.
Therefore, the dynamics of the system can be simplified to four
ordinary differential equations for the evolution of the free contour
lengths $l_a(t)$ and $l_d(t)$, respectively, the number of attached
filaments $n_ a(t)$ (the total number of filaments $N$ is constant and
therefore $n_d(t)=N-n_a(t)$) and the distance $\xi(t)$ between the obstacle
and the grafting point.

The solution behavior of this system of equations has been analyzed
numerically depending on the maximal grafting speed $v_g^{max}$ and
the rate constant for the attachment of filaments $k_a$. Fortunately,
most other model parameters are known experimentally.
The model displays two different dynamical regimes: steady and
oscillatory motion, whereby the oscillatory regime is robust against
changes in the parameters. The oscillations occur on the time scale of
min and produce jumps of the obstacle displacement in the \textmu{}m
range and resemble very much relaxation oscillations.

A deeper analysis of the solutions suggests, that the oscillations
arise from a so called push-pull mechanism, i.e.\ a competition
between pulling and pushing forces acting on the obstacle.
In this mechanism, a long pull phase, where most of the filaments are
attached to the obstacle and polymerization stalls, alternates with a
short push phase, where most filaments are detached and polymerize
rapidly.
In the pull phase the grafting velocity $v_g$ is higher than the
polymerization speed $v_p$, and the magnitude of the forces on the
obstacle increases due to their dependence on $\xi$, $l_a$, and $l_d$.
At a certain point, in the pushing phase, the pushing forces
outweigh the pulling forces and cause an avalanche like detachment
of filaments and the obstacle ``hops'' forward, lowering the load on
the detached filaments, which start to polymerize rapidly and thus
$v_g<v_p$.
Meanwhile free filaments start to attach to the obstacle and increase
the pulling force, i.e.\ the obstacle slows down.  Free filaments start
to buckle, the polymerization stalls and the cycle reenters the
pulling phase.

Obviously, this complex cycle arises from the subtle interplay between
pushing and pulling forces on the one hand and the grafting and
polymerization velocities on the other hand. It would be an
interesting task, to explore the parameter space a little bit further
and identify the absolutely necessary ingredients to find oscillations
in the push-pull mechanism.

Besides steady and oscillatory motion the model also yields bistable
and excitable behavior, which might lead to the reinterpretation of
previous experiments and is reminiscent of the behavior caused by
nonlinear friction in a variety of systems with a complex surface
chemistry \cite{UKG04}.
Another aspect for future work is to couple the microscopic dynamics of the
filament brush to the bulk mechanics of the cross-linked gel.
%
%
\subsection{Macroscopic models}
While microscopic models give a detailed description of the
polymerization and cross-linking dynamics, macroscopic models are
more concerned about the global stress distribution in the gel, but
adopt a more general formulation for the interface dynamics. In the
following we will shortly introduce and discuss the essence of three
simple models, i.e.\ by Lee et al. \cite{LLK05}, Sekimoto et
al. \cite{SPJ04}, and John et al. \cite{JPK08}, describing the
symmetry-breaking in an actin gel around a solid bead. We will
conclude this section with the discussion of a more advanced
mechanical model using homogenization techniques proposed by Caillerie
et al. \cite{CJM09}.
\subsubsection{A phenomenological model of symmetry-breaking}
As an introduction we will present the problem of
symmetry-breaking from a purely phenomenological point of view as
proposed by Lee et al. \cite{LLK05}. If a bead moves, it is natural
to assume that a force is applied on the bead, probably due to
deformed filaments that releases stress on the bead.
Let $g({\bf r},t)$ denote the force per unit area in the normal
direction that is exerted on the bead, which could be e.g.\ a function
of the local actin concentration. The total force on the bead is given
by ${\mathbf F} =\int g({\bf r},t) {\mathbf n} dA$, with ${\mathbf n}$
being the unit normal vector on the bead.
The velocity of the bead is related to the total force via a linear
relation
\begin{equation}
{\mathbf v}= \xi {\mathbf F} =\xi \int g({\bf r},t) {\mathbf n} dA\,,
\label{vmotion}
\end{equation}
where $\xi$ is a dissipative coefficient, taken to be scalar for
simplicity (it is necessarily so for a sphere in a Newtonian fluid).

It is then assumed that the rate of change of $g({\bf r},t)$ is a
local function of the bead velocity ${\bf v}$, and that there is a
feed back of the motion on the force $g$: faster motion is associated
with a decrease of polymerization in the front and an increase of
polymerization at the rear, and hence has an impact on $g$.
Under the assumption of analyticity, the evolution of $g$ can be written
as
\begin{equation}
\partial_t g = - g - g^2- c g^3 + a{\mathbf v}.{\mathbf n}+ b g
{\mathbf v}.{\mathbf n} \label{gmotion}\,,
\end{equation}
with $c>0$ in order to ensure stability of the homogeneous stationary
state $g=0$. The signs of $a$ and $b$ are left arbitrary for the
moment.
Eqs.\ (\ref{gmotion}) and (\ref{vmotion}) constitute a complete set
that can be solved numerically or analytically using a perturbation
ansatz as will be outlined in the following.

\begin{figure}[hbt]
\begin{center}
\includegraphics[width=0.7\hsize]{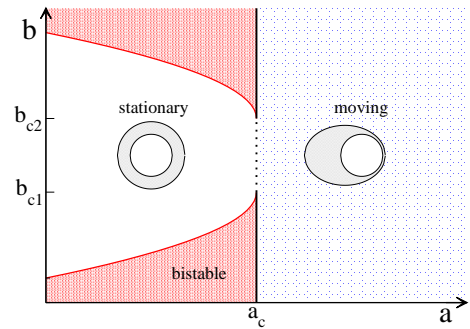}
\end{center}
\caption{Phase diagram arising from the analysis of Eq.\
(\ref{gmotionp}).  The transition between the stationary (white
region) and moving (light shaded region) state occurs at
$a=a_c$. For $b_{c1}<b<b_{c2}$ the bifurcation is supercritical. In
the dark shaded region there coexist two solutions ${\mathbf v}=0$ and
${\mathbf v}=const\ne0$ which are linearly stable. Adapted from Ref.\
\cite{LLK05}.\label{figkardar}}
\end{figure}

It is a simple matter to see that the set [Eqs.\ (\ref{gmotion}) and
(\ref{vmotion})] admits the fixed point $g={\mathbf v}=0$. By
superposing small perturbations on this solution, reporting into the
above set, and expanding up to linear order in the perturbations, one
finds that the fixed point is unstable for $a>3/\xi\equiv a_c$ and
stable otherwise.
If motion takes place then this means that a symmetry-breaking has
occurred, and thus the local force $g$ has lost the spherical
symmetry.  More precisely by expanding $g$ in spherical harmonics,
$g=\sum _{\ell m} g_{\ell m} Y_{\ell m} (\theta,\phi)$, and reporting
into (\ref{gmotion}) by assuming a direction of motion, say along
$oz$, one finds to leading order
\begin{equation}
\partial_t g_{\ell m} = \left[1-  {a\over a_c}\delta_{\ell 1}\right] g_{\ell m}+\ldots
 \label{gmotion0}
\end{equation}
whereby the ``$\ldots$'' refer to nonlinear terms.
Integration of other harmonics than the mode $\ell=1$ in
(\ref{vmotion}) vanishes exactly due to symmetry.
Beyond the symmetry-breaking bifurcation for $a>3/\xi$ the
perturbations grow exponentially in time, and nonlinear terms are
needed.
Since only the mode $\ell=1$ is excited, the other modes are treated
as adiabatically enslaved to it (at least in the vicinity of the
threshold).
The following analysis consists in expanding the solution for higher
order terms and expressing the amplitudes of the higher harmonics in
terms of the amplitude of the first order mode. Once $g$ has been
replaced by ${\mathbf v}$ from (\ref{vmotion}) and inserted into
(\ref{gmotion}) a closed equation for ${\mathbf v}$ (which is simpler
to assess experimentally than $g$) is obtained \cite{LLK05}
\begin{equation}
\partial_t {\mathbf v}  = \epsilon {\mathbf v} +{27\over 5\xi^2}\left[ \left({\xi b\over 3} -1 \right ) +\left ({\xi b\over 3} -2 \right )-c\right] {\mathbf v}^3
+ u {\mathbf v} ^5 \,.
 \label{gmotionp}
\end{equation}
$\epsilon \equiv (1- {a/ a_c})$ is a small parameter (expressing the
fact that we focus on the instability threshold), and we have used the
convention ${\mathbf v}^3=v^2{\mathbf v} $, and so on. The expression
of $u$ in terms of the coefficients $a$, $b$, $c$, and $\xi$ is not
shown here. However, it is reported in \cite{LLK05} that $u$ is
negative in the considered parameter space.
The amplitude equation (\ref{gmotionp}) serves to discuss the
phenomenology of the motion, that we will briefly summarize here and
which is shown schematically for the $b-a-$ parameter plane in Fig.\
\ref{figkardar}.

If the coefficient of the cubic term is negative, then a continuous
transition from the symmetric (motionless) to the asymmetric (moving)
state occurs at $\epsilon=0$, i.e.\ the bifurcation is supercritical
(an analogue of a second order transition). This happens for a certain
range of the product $\xi b$ and by fixing $c$ to unity for
definiteness.
For a certain range of the product $\xi b$
the cubic coefficient changes sign, a signature of a subcritical
bifurcation (an analogue of a first order transition). This is known
to lead to multistability: in a certain parameter range for
$\epsilon<0$ there coexist two solutions: ${\mathbf v}=0$ and
${\mathbf v}=const$, both solutions being locally stable, i.e.\ stable
with respect to small perturbations. One may think that the system may
switch back and forth between the two solutions.

Several remarks are in order.
(i) The above discussion is fully
phenomenological, and actually Eq.\ (\ref{gmotionp}) could have been
written directly. However, the derivation given in \cite{LLK05} has a
certain merit with regard to the description of the general laws
behind the motion, and the feedback mechanism.
(ii) The coefficients
are certainly complicated functions of experimental parameters, and a
microscopic model is needed in order to relate phenomenological and
experimental parameters.
(iii) The origin of the forces acting on the bead and the subsequent
generation of motion in this model is not clear. We will come back to
this vital question in the final conclusions.
\subsubsection{The role of tensile stress during the symmetry-breaking in actin gels}
\begin{figure}
\begin{center}
\includegraphics[width=0.7\hsize]{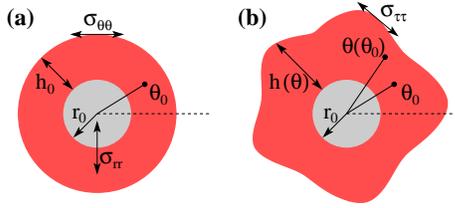}
\end{center}
\caption{Schematic view of a bead surrounded by an elastic gel,
  showing several definitions as explained in the text. Shown is a
  symmetric gel (left) and an asymmetric gel (right). In the latter
  case the tensile stress is rather generally denoted by
  $\sigma_{\tau\tau}$ with $\tau$ designating the tangent direction.
  \label{scheme1}}
\end{figure}

%
%
A first physical macroscopic model including elasticity has been put
forward by Sekimoto et al.  \cite{SPJ04} in order to explain the birth
of the symmetry-breaking of an initially symmetric gel layer growing
on a spherical or cylindrical object. The crux of their analysis is
that the gel that has been formed is being continuously pushed
outwards due to the arrival of new monomers at the bead. This is
supposed to lead to large lateral stresses on the outer gel
interface. A schematic representation of the model is shown in Fig.\
\ref{scheme1}.
In the following we outline the model in a cylindrical geometry. If
$\sigma_{\theta\theta}$ designates the tangential stress, and if the
gel layer is axisymmetric, it is postulated that
\begin{equation}
\sigma_{\theta\theta}=E {r-r_0\over r_0}\,, \label{eqsek1}
\end{equation}
where $E$ is the Young modulus. At $r=r_0$ (the bead radius) the
tensile stress vanishes, expressing the assumption that the gel layers
are added unstretched at the bead surface.
If $h$ denotes the total gel thickness, one can write for the tensile
stress at the external gel surface $\sigma_{\theta\theta}|_{r=r_0+h}=E
{h\over r_0}$. The external surface of the gel is not subject to a
force, so that both the tangential and normal forces must vanish, i.e.\
$\sigma_{r\theta}|_{r=r_0+h}=\sigma_{rr}|_{r=r_0+h}=0$.
The tangential force is also zero at the bead surface,
$\sigma_{r\theta}|_{r=r_0}$. For a symmetric gel the shear stress
vanishes everywhere in the gel.  If a non zero tensile stress exists
[Eq.\ (\ref{eqsek1})], the mechanical equilibrium in the bulk, i.e.\
\begin{equation}
div(\sigma)=0 \label{divsigma}
\end{equation}
provides the following relation between the tensile and radial stress
distribution, $\sigma_{\theta\theta}$ and $\sigma_{rr}$,
\begin{equation}
\sigma_{rr}= \frac{E}{2r_0r}\left[(r-r_0)^2-h^2\right]\,.
\end{equation}
The gel may grow because of a gain in polymerization energy at the
bead surface.
This growth takes place at a certain price: the higher the thickness
of the gel is, the larger is the stored elastic energy. Therefore, one
expects the growth to stop at a certain equilibrium thickness
$h_0$. The following growth kinetic relation has been suggested
\cite{SPJ04}
\begin{equation}
\partial_t h= k_pe^{c_p\sigma_{rr}|_{r=r_0} } -k_d
e^{c_d\sigma_{\theta\theta}|_{r=r_0+h}} \label{dhdt}
\end{equation}
 where $k_p$, $k_d$, $c_p$, $c_d$ are positive constants.  The first
term, which is positive, accounts for the polymerization at the bead,
while the second one, which is negative, refers to depolymerization at
the external surface. Note that on the one hand
$\sigma_{rr}|_{r=r_0}=-Eh^2/(2r_0^2)<0$, and thus stress penalizes
polymerization and acts against gain in chemical bonds at the bead. On
the other hand $\sigma_{\theta\theta}|_{r=r_0+h}>0$ and this causes
the depolymerization to increase with the gel thickness, since
$\sigma_{\theta\theta}|_{r=r_0+h}$ increases with $h$.
Setting $\partial_t h=0$ provides us with the steady state thickness
$h_0$ as a function of other parameters. One straightforwardly finds
\begin{equation}
{c_p\over c_d} ({h_0\over r_0})^2 + {h_0\over r_0}-{2\over c_d
E}\ln({k_p\over k_d})=0\,.
\end{equation}
A steady solution exists as long as $k_p/k_d>1$. It can easily be
checked that the steady solution $h_0$ is stable with respect to a
homogeneous, i.e.\ axisymmetric, increase or decrease of $h_0$. One
further sees that $h_0$ is a linear function of $r_0$, since only the
ratio $h_0/r_0$ enters the above equation.  This seems to be in good
agreement with experimental observations \cite{NGF00}.

Let us now discuss the linear stability analysis of the steady
solution with respect to perturbations that break the circular
symmetry. In a cylindrical geometry the eigenmodes are $\sim e^
{im\theta_0}$ (or $\cos(m\theta_0)$ in real variables) where
$\theta_0$ is the angular variable shown in Fig.\ \ref{scheme1}, and
$m$ is an integer. Because the equations are autonomous with respect
to time, all the eigenmodes can be written as $\sim e^{\beta t}$,
where $\beta$ is the amplification or attenuation rate of the
perturbation that must be determined from the model equations.
We set for the perturbed thickness
\begin{equation}
h(\theta_0)= h_0 [1+\epsilon_m(t) \cos(m\theta_0 )]\,. \label{perturb}
\end{equation}
$\epsilon_m$ is a small, time-dependent quantity that justifies the
linear stability analysis. The stress field inside the gel will thus
be parametrized by the function $h(\theta_0 )$.
The linear analysis now consists in (i) solving the stress field in
the gel with the appropriate boundary conditions, and (ii) using
the kinetic relation (\ref{dhdt}) to obtain the dispersion relation
$\beta=f(m,{p})$, where ${p}$ is an abbreviation for all the physical
and geometrical parameters that enter into the equation.

In order to solve for the elastic field, one needs to specify a
constitutive law. Unlike in the two models \cite{CJM09,JPK08} we
will discuss in the following sections, where a constitutive law is
used by evoking basic continuum mechanics concepts, the Sekimoto et
al.\,model \cite{SPJ04} is based on an extension of the postulate
represented by Eq.\ (\ref{eqsek1}) to a modulated thickness
$h(\theta_0)$. The basic ingredient of their analysis is to
introduce an unknown function $\theta (\theta_0 )$ such that a
material point of the gel which is originally located at $\theta_0$
in an undeformed reference state is moved to a new position
$\theta(\theta_0)$ upon deformation. Then the elongation ratio
$(r-r_0)/r_0$ is replaced by $(rd\theta(\theta_0) -
r_0d\theta_0)/(r_0d\theta_0)$ so that the tensile stress takes the
form
\begin{equation}
\sigma_{\theta\theta}=E \left[{r\over r_0} {d\theta(\theta_0 )\over
d\theta_0} -1\right ]\,. \label{cons1}
\end{equation}
By using the equilibrium balance condition Eq.\ (\ref{divsigma}) and
neglecting the shear stress (for a discussion of this assumption see
Ref.\ \cite{SPJ04}) one can again find a relation between $\sigma_{rr}$
and $\sigma_{\theta\theta}$.
By defining
\begin{equation}
T\equiv \int _{r_0} ^ {r_0+h}\sigma_{\theta\theta} dr \label{Tdef}
\end{equation}
one deduces that
\begin{equation}
\sigma_{rr}|_{r=r_0}=-\frac{T}{r_0}\,,
\end{equation}
since only this quantity enters the kinetic relation (\ref{dhdt}) that
is needed for the derivation of the dispersion relation. Recall also,
that $\sigma_{rr}|_{r=r_0+h}=0$ to linear order in the perturbation
$\epsilon_m$.
Upon using (\ref{cons1}) the integral (\ref{Tdef}) leads to
\begin{equation}
T=  E \left[ \left( h(\theta_0)+ {h(\theta_0)^2\over 2 r_0}\right)
{d\theta(\theta_0 )\over d\theta_0} -h(\theta_0)\right]\label{Tdef1}
\end{equation}
Additionally, Eq.\ (\ref{divsigma}) yields, under the assumption of
zero shear stress, $\partial_\theta T=0$, that is $T$ is independent
of $\theta$.
This condition provides a relation between $\theta(\theta_0)$ and
$h(\theta_0)$. Upon using $\int _0^{2\pi} {d\theta(\theta_0 )\over
d\theta_0} d \theta_0=2\pi$, one can express $T$ as a function of an
integral $\int _0^{2\pi} F(h(\theta_0))\,d\theta_0$ where only
$h(\theta_0)$ enters ($F$ is given by Eq.\ (D5) of Ref. \cite{SPJ04}).
Plugging this relation into (\ref{Tdef1}) provides a relation between
$d\theta/d\theta_0$ and $h(\theta_0 )$, and substituting $h$ by Eq.\
(\ref{perturb}) $\sigma_{\theta\theta}|_{r=r_0+h}$ and $\sigma_{rr}
|_{r=r_0}$ can be deduced to first order in $\epsilon_m$.
After some algebraic manipulations the dispersion relation is obtained
\begin{equation}
\beta = {\Omega_m k_d\over r_0}\,,
\end{equation}
where
\begin{equation}
\Omega_m =c_d E e^{c_dE \bar h_0} {\bar h_0\over \bar h_0+2}
\end{equation}
and where we have set $h_0/r_0\equiv \bar h_0$. $\beta$ is positive,
meaning that the perturbation grows exponentially with time: the
symmetric gel layer is thus unstable. Surprisingly, the dispersion
relation does not depend on $m$ (since $\Omega_m$ does
not). Consequently, all wavenumbers have the same growth rate. Thus,
the linear stability analysis does not select a typical mode $m$ (like
the fastest growing mode) for the instability.
\subsubsection{A nonlinear study on symmetry-breaking in actin gels}
Unlike the previous model where a tensile stress distribution is {\it
a priori} postulated, the idea of the model proposed by John et
al. \cite{JPK08} is to treat the actin gel as an elastic continuum in
the framework of a linear theory, and to formulate a simple kinetic
relation expressing growth (or polymerization), different from Eq.\
(\ref{dhdt}).

\begin{figure}
\begin{center}
\includegraphics[width=0.35\hsize]{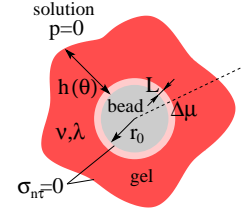}
\end{center}
\caption{Schematic view of a bead surrounded by an elastic gel,
  showing also several definitions of Ref.\
  \cite{JPK08} as explained in the text. \label{scheme2}}
\end{figure}

The model considers a bead (radius $r_0$) surrounded by a growing
elastic actin gel (radius $r_0+h$) as shown in Fig.\ \ref{scheme2}.
The gel is stressed by a small molecular displacement $L$ in normal
direction at the bead/gel interface, i.e.\
\begin{equation}
u_r|_{r=r_0}=L \label{bc1}\,,
\end{equation}
where $u_r$ denotes the radial component of the displacement.
This choice is motivated by the microscopic picture, that for the
addition of monomers, enzymes facilitate a molecular displacement $L$
at the bead/gel interface. This displacement is the source of stress.
The bead/gel interface as well as the external gel surface is shear
free and the normal stress at the external surface is set to
zero\footnote{Actually it can be set to $-p$, where $p$ denotes the
liquid pressure, but this contribution is quite small.}.
No condition is imposed on the normal
stress component at the bead, since there a displacement is imposed
instead. The boundary conditions on the stress are thus
\begin{eqnarray}
\sigma_{n\tau} & \equiv & n_i\sigma_{ij} \tau_j|_{r=r_0, r=r_0+h}=0\nonumber\\
\sigma_{nn} & \equiv & n_i\sigma_{ij} n_j|_{r=r_0+h}=0\,, \label{bc2}
\end{eqnarray}
where $n_i$ and $\tau_i$ are the $i$th component of the unit normal
and tangent vector of the surface under consideration (bead or
external gel surface).

The stress distribution in the gel is obtained then by solving the
Lam\'e equation for the displacement field
\begin{equation}
\nabla^{2}{\mathbf u}+\frac{1}{1-2s}{\mathbf \nabla}({\mathbf
\nabla}.{\mathbf u})=0\,, \label{lame}
\end{equation}
where $s$ is the Poisson ratio. The stress is related to the
displacement ${\bf u}$  by Hooke's law
\begin{equation}
\sigma_{ij}=2\nu \epsilon_{ij}+\lambda\epsilon_{kk}\delta_{ij}\,,
\label{hooke}
\end{equation}
where $\epsilon_{ij}=(\partial_i u_j + \partial_j u_i)/2$ is the
strain tensor, and $\lambda$ and $\nu$ are the Lam\'e coefficients
which are related to the Young modulus $E$ and $s$ (for an isotropic
material there are only two independent elastic parameters).

Eqs.\ (\ref{bc1})\,-\,(\ref{lame}) represent a complete set that
allows to determine, the stress and displacement fields in the
gel. Note that, despite the fact that the bulk equations are linear,
the problem acquires a nonlinear character via the geometry of the
external gel boundary.
Indeed if we fix a certain
arbitrary geometry $h(\theta)$, then the stress and displacement will
be a nonlinear function of $h$. The calculation can be handled
analytically for a symmetric gel as well as in the linear stability
analysis \cite{JPK08}. Beyond a linear analysis, a numerical study has
been performed and will be briefly discussed below.

Once the mechanical problem is solved, one needs to compute the cost
in elastic energy per unit mass for inserting a monomer on the
bead/gel interface, i.e.\ the chemical potential difference.
For the sake of simplicity we focus on the case of an inert external
gel surface, i.e.\ neither polymerization nor depolymerization takes
place at the external interface.
We assume that insertion of a monomer at the bead results in a
displacement of the external gel surface in the radial direction
proportional to the chemical potential change at the bead.
We will critically assess this assumption in the next section on
homogenization models.

One may then write a kinetic relation of the shape evolution of the
gel envelope
\begin{equation}
\partial_t h = -M\Delta\mu \label{dhdtp}\,,
\end{equation}
where $M$  denotes a mobility and $\Delta\mu$ the difference in the
chemical potential between a volume element in the gel and in
solution at the internal  interface.
Here we assume that the mobility is associated with the
polymerization/depolymerization kinetics, which constitutes the
prevailing dissipation mechanism. The chemical potential is composed
of a contribution due to the gain in polymerization (denoted as
$\Delta\mu_p<0$) and an elastic part \cite{JPK08}
\begin{equation}
\Delta\mu = \Delta\mu_p+  \nu u_{ij}u_{ij}+\frac{\lambda}{2}
u^2_{kk} -\sigma_{nn}(1+u_{kk}) \label{dmu}
\end{equation}
Note that (\ref{dhdtp}) differs from (\ref{dhdt}) not only by the
presence of the exponential function (which can be linearized since
the stress energy is always small in comparison to the thermal
excitation energy; hidden in the constant $c_p$), but most importantly
by the stress combination. Eq.\ (\ref{dhdt}) contains only a linear
form and no quadratic forms as in Eq.\ (\ref{dmu}). Here, the
quadratic form is essential for the mode selection leading to a comet
formation, as will be described below.

The stress problem Eqs.\ (\ref{bc1})-(\ref{lame}) can easily be solved
analytically for a spherical geometry (axisymmetric growth). Upon
setting $\partial_t h=0$ in Eq.\ (\ref{dhdtp}) one finds a steady
solution with a gel thickness $h_0$ obeying \cite{JPK08}
\begin{equation}
h_0=\left[\left(2\frac{E \alpha - (1-2s )\Delta\mu_p}{ 2 E \alpha
+(1+s)\Delta\mu_p}\right)^{1/3}-1 \right]r_0\,,
\end{equation}
where  $\alpha=L/r_0$. This solution exists for $2 E
\alpha/(1+s) \ge -\Delta\mu_p$: elasticity  acts against
monomer addition, so that the gel stops growing at that thickness.
In the opposite limit growth continues without bound. Both situations
have been observed experimentally \cite{PLP04}, however in the latter
case growth stopped due to monomer depletion at the bead/gel
interface.
%

\begin{figure}
\begin{center}
\includegraphics[width=0.5\hsize]{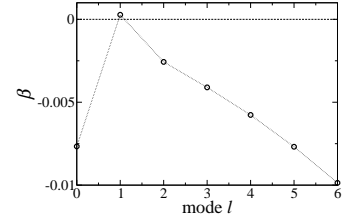}
\end{center}
\caption{The dispersion relation as a function of the mode $\ell$
obtained from the linear stability analysis of the John et al.  model
\cite{JPK08}. The fastest growing mode corresponds to $\ell=1$.
  \label{dispJohn}}
\end{figure}

\begin{figure}[hbt]
\begin{center}
\includegraphics[width=0.45\hsize]{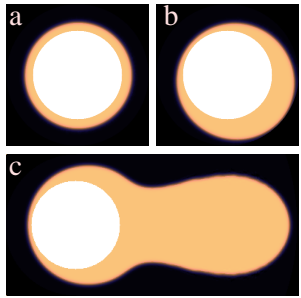}
\end{center}
\caption{ Symmetry-breaking of a circular gel. Shown is the evolution
of the gel thickness, starting from a homogeneous thin gel (a) with
random small amplitude perturbations.  (b) shows the initial symmetry
breaking, while (c) shows the subsequent evolution of the shape into a
comet in the far nonlinear regime (adapted from Ref.\
\cite{JPK08}).\label{fig_cometP}}
\end{figure}

The linear stability analysis of the symmetric case can be performed
analytically (by decomposing the stress and the shape evolution into
spherical harmonics $Y_{\ell m}$). The dispersion relation $\beta(\ell)$ is
presented in Fig.\ \ref{dispJohn}.
The basic result is that a symmetric shape is unstable against
symmetry breaking \cite{JPK08}. Interestingly the mode which
corresponds to a translation of the external surface with respect to
the bead is the most unstable\footnote{We should not confuse the mode
$\ell=1$ with the usual global translation, which is a neutral
mode. Here only the external gel surface moves while the bead is
fixed, so that the mode $\ell=1$ is a physical one.}. This translation
motion is similar to that shown in Fig.\ \ref{fig_cometP}\ (b).
For this instability the quadratic terms in Eq.\ (\ref{dmu}) play a crucial role,
since considering only the linear terms leads to a stable symmetric
solution, with a zero growth rate for the translational mode.
%

In order to ascertain the subsequent evolution of the external
boundary (i.e.\ in the fully nonlinear regime), a full numerical
analysis has been performed \cite{JPK08}.
The set of mechanical equations (\ref{bc1})\,-\,(\ref{lame}), and the
growth kinetics (\ref{dhdtp}) have been cast into a phase-field
approach, which has now become a frequent method to treat free moving
boundary problems.
For the details of the phase-field formulation and their numerical
implementation we would like to refer the reader to the original paper
\cite{JPK08}. Here we shall only report the basic results

The numerical study in two dimensions with plane strain shows that for
axisymmetric initial conditions with small amplitude perturbations,
the symmetry is broken for the mode $m=1$ (identical to the mode
$\ell=1$ in three dimensions), which corresponds to a translation of
the gel layer with respect to the bead.
Where this symmetry-breaking occurs, depends only on the initial
conditions.  The instability then evolves further into an actin comet,
reminiscent of the comet developed by {\it Listeria monocytogenes}.
Fig.\ \ref{fig_cometP} shows a typical result of a numerical
simulation.
The comet formation seems here to be the generic growth
mode. This finding points to the fact that the comet formation is
probably a quite robust feature; it results from simple physical
prototypes.

The physical picture of the symmetry breaking may be understood as
follows. Let us start with a symmetric layer as in Fig.\
\ref{fig_cometP}\ (a). Suppose that due to some natural (inevitable)
fluctuation, the gel layer becomes asymmetric, as in Fig.\
\ref{fig_cometP}\ (b). The stress is due to addition of new monomers
at the bead/gel interface.
On the side where the gel thickness is small the stress field is
stronger than on the other side, since the gel feels more the
``outwards pushing'' of new monomers inserted at the bead. Because of
the increase of the stress (and strain) in the thinner layer,
polymerization becomes unfavorable there. New monomers will
preferentially be inserted on the side where the thickness is larger.
The appearance of modes larger than $\ell=1$ would create several thin
and thick regions, which are likely unfavorable. It seems thus that the
mode $\ell=1$ is optimal for the insertion of monomers at the bead/gel
interface.

By considering also the stress dependent depolymerization at the
external gel interface, one finds another instability, whereby the
location of the most unstable mode is determined by surface tension.
This result is in agreement with the occasional experimental
observation of higher order modes \cite{DSR08}, which will be
discussed next within a more elaborate homogenization model.
\subsubsection{Homogenization models}
So far, continuous models have been suffering from the inadequate or
insufficient description of the mechanical aspect of the actin gel.
The greatest difficulty in the description of the mechanical
equilibrium of the growing actin network arises from the fact, that
the growth history determines the network structure and therefore also
the stress distribution.
Consequently, a realistic model would have to include the information
on how the network evolved.

A second problem, which is most prominent in the description of the
symmetry-breaking around solid objects, is the coupling of the
growth process between the internal and the external interface.
As an example, consider the problem of actin growth around a bead
functionalized with ActA or Wasp/Scar: Typically growth takes place at
the barbed end of the actin filaments, which points toward the bead/gel
interface. This growth process pushes older gel layers further away
from that interface.
Experimentally, the growth process is observed by an increase in the
gel thickness. However, it is not clear how the insertion of mass at
a (fixed) solid/gel interface translates into the displacement of the
(free) gel/liquid interface.
A realistic elastic theory would have to account for this coupling
problem in a rigorous way.
%

One observation, which might help to solve the above mentioned
problems at least partly, is that the actin network forms more or less
regular structures, which are not perfectly periodic, but could be
considered in a first approximation as ``almost periodic''.
The actin gel can then be regarded abstractly as a network of elastic
filaments connected by nodes with a certain periodicity. This network
is completely defined by the positions of the nodes and their
connectivities.
In the network structure, the size of each elementary cell, e.g.\ the
distance between two Arp2/3 crosslinks ($\sim$ several tens nm) , is
small compared to the total size of the structure
($\sim$1\,\textmu{}m), which introduces a small parameter $\eta$ into
the problem, which is the ratio of the length of the unit cell and the
total network size.
In the following we will shortly outline the basic idea of an actin
homogenization model in two dimensions \cite{CJM09}.

\begin{figure}
\begin{center}
\includegraphics[width=0.6\hsize]{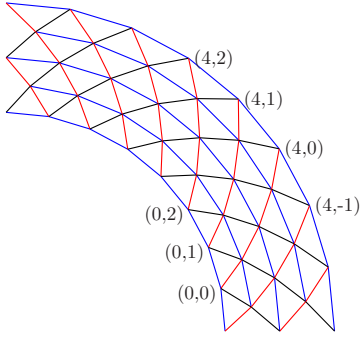}
\end{center}
\caption{Sketch of a small part of the filament network with some
examples for the node numbering ($\nu^1$,$\nu^2$). The ``curves''
along which $\nu^1$=const. and $\nu^2$=const. are shown in blue and red,
respectively.
\label{nwtop}}
\end{figure}

We consider a planar network of stiff elastic bars around a solid
cylinder (radius $r_0$) with the topology shown in Fig.\ \ref{nwtop}.
The bars are connected to each other via nodes.
The actin filaments are assumed to be linked to the cylinder at
$N_{t}$ sites evenly located on its surface at a distance $p=\eta r_0$
between two close sites, i.e.\ at an angular distance $\eta =\frac{2\pi
}{N_t}$. The actin gel is made of $N_{n}$ layers of bars in the radial
direction.
As the growth of the gel is due to the polymerization of actin
monomers at the surface of the cylinder, each layer is assumed to be
made up of the same number of nodes. So the nodes of the gel can be
numbered by two integers $\left({\nu ^{1},\nu ^{2}}\right)$ with
$\nu^{1}$ numbering the radial layers and $\nu^{2}$ the position of
the node in each layer, respectively.

It is assumed that the discrete net is made up of a large number of
bars meaning that $N_{t}$ and $N_{n}$ are very large and of the same
order. To be more precise, the parameter $\eta $ is assumed to be very
small and the number $N_{n}$ of layers is given by $N_{n}=\frac{\alpha
}{\eta }$ with $\alpha $ being of order 1 with respect to $\eta $.
Using this notation, a node of the gel can be labeled by $\left({\mu
^{1\eta },\mu ^{2\eta }}\right)$ with $\mu ^{i\eta }=\eta \nu ^{i}$.
The coordinates $\left({\mu ^{1\eta },\mu ^{2\eta }}\right)$ take
values in $\omega =]0,\alpha [\times]0,2\pi [$ and are meant to
become the set of Lagrangian curvilinear coordinates of the equivalent
continuous medium.

\begin{figure}
\begin{center}
\includegraphics[width=0.7\hsize]{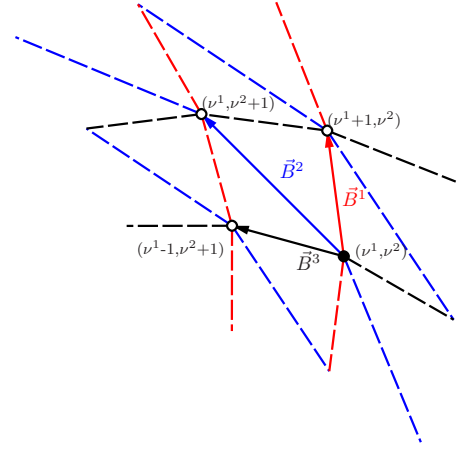}
\end{center}
\caption{\label{unitcell}Sketch of an elementary cell, showing the
node numbering $(\nu^1,\nu^2)$ and the elementary bar vectors $\vec
B^b$ (solid arrows). All other bars are shown as dashed lines. The
``curves'' along which $\nu^1$=const. and $\nu^2$=const. are shown in
blue and red, respectively.}
\end{figure}

The upscaling of the net to a continuous medium consists in
determining the equivalent stresses from the bar tensions, the
equations of equilibrium (or motion) satisfied by these stresses and
an equivalent constitutive equation ensuing from the properties of
the bars.
This can be carried out by using an asymptotic expansion (for an
introduction see \cite{CaC01,Lov44,ToC98}). Here, as the network
structure is simple, a more heuristic presentation can be used.
The basic idea of the homogenization process is that, for most of the
motions of the network, the positions of its nodes $\left({\nu
^{1},\nu ^{2}}\right)$ can be approximated by a continuous function
$\vec{\psi }\left({\nu ^{1},\nu ^{2}}\right)$ where $\mu ^{i\eta
}=\eta \nu ^{i}$.
The purpose then is to determine the equations governing this
deformation function.  The equivalent Cauchy stress tensor is given by
the classical relation due to Cauchy \cite{CaC01,Lov44}
\begin{equation}  \label{stress1}
\sigma =\frac{1}{g}   \sum\limits_{\mit b = 1}^{3}N^{b}\vec{e}\,^{b}\otimes \vec{B}^{b}
\end{equation}
with $\vec{B}^{b}$ being the "bar vector" linking the two extremities
of the bar $b$ ($b=1,2,3$) of the elementary cell (shown in Fig.\
\ref{unitcell}) $\left({\nu ^{1},\nu ^{2}}\right)$ of the network in a
deformed state,
$\vec{e}\,^{b}=\frac{\vec{B}^{b}}{\vert{\vert{\vec{B}^{b}}|}|}$ being
the corresponding unit vector, $N^{b}$ the tension in the bar and
$g=\left|\!\left|{\vec{B}}^{1}\wedge {\vec{\mit B}}^{\mit
3}\right|\!\right|$ being the surface of the elementary cell.
The constitutive equation of the equivalent continuous medium follows
from the constitutive equations of the bars which, for the sake of
simplicity, are assumed to be
%
 \begin{equation}
N^{b}=k^{b}\frac{l^{b}-l_{m}^{b}}{l_{m}^{b}}
\end{equation}
with $l^{b}=\vert{\vert{\vec{B}^{b}}|}|$.
$l_m^{b}$ is the length of the bar $b$ at rest and serves as a
parameter in the constitutive equation.

In order for a symmetrical equilibrium configuration of the gel to be
possible, the constitutive equations of the bars 1 and 3 should be
identical, i.e.\ $k^{1}=k^{3}$ and $l_{m}^{1}=l_{m}^{3}$.

Following the homogenization assumption stating that the position of
the node $\left({\nu ^{1},\nu ^{2}}\right)$ is $\vec{\psi }\left({\nu
^{1},\nu ^{2}}\right)$ with $\mu ^{i\eta }=\eta \nu ^{i}$, a simple
Taylor expansion yields:
%
 \begin{eqnarray}
\vec{B}^{1} & = &\eta \left({ \partial _{1}^{\mu }\vec{\psi }+\partial _{2}^{\mu}\vec{\psi}}\right)\nonumber\\
\vec{B}^{2} & = & \eta \partial _{2}^{\mu }\vec{\psi }\\
\vec{B}^{3} & = &\eta \left({-\partial _{1}^{\mu }\vec{\psi }+\partial _{2}^{\mu }\vec{\psi }}\right)\nonumber
\end{eqnarray}
where $\partial _{i}^{\mu }=\frac{\partial }{\partial \mu ^{i}}$.
Since they are associated with a quite simple numbering system for the
nodes, the variables $\mu ^{1}$ and $\mu ^{2}$ arise naturally as
Lagrangian variables of the equivalent continuous medium through the
homogenization process. However, they are not the most convenient
variables to study symmetrical equilibrium configurations. Therefore
we have introduced the variables $\lambda ^{1}$ and $\lambda ^{2}$
defined by
%
  \begin{equation}
\lambda ^{1}=\mu ^{1}\,\textrm{and}\,\lambda ^{2}=\frac{\mu ^{1}}{2}+\mu ^{2}\,.
\end{equation}
Setting
%
\begin{equation}
\vec{\varphi }\left({\lambda ^{1},\lambda ^{2}}\right)=\vec{\psi
}\left({\lambda ^{1},-\frac{\lambda ^{1}}{2}+\lambda ^{2}}\right)
\end{equation}
one finds
%
\begin{equation}
\vec{B}^{1}=\eta \vec{g}_{1} ,\, \vec{B}^{2}=\eta \vec{g}_{2} \textrm{
and } \vec{B}^{3}=\eta
\left({-\vec{g}_{1}+\frac{1}{2}\vec{g}_{2}}\right)
\end{equation}
with $\vec{g}_{1}=\partial _{1}^{\lambda }\vec{\varphi }$ and
$\vec{g}_{2}=\partial _{2}^{\lambda }\vec{\varphi }$.

Carrying these relations into (\ref {stress1}) yields
\begin{equation}
\sigma=\frac{1}{\vert\vert\vec{g}_1\wedge\vec{g}_2\vert\vert}\sum_{i=1}^3\vec{S}^i\otimes\vec{g}_i
\end{equation}
with $\vec{S}^{1}=\eta
\left({N^{1}\vec{e}\,^{1}-N^{2}\vec{e}\,^{2}}\right)$ and
$\vec{S}^{2}=\eta
\left({\frac{1}{2}N^{1}\vec{e}\,^{1}+N^{2}\vec{e}\,^{2}+\frac{1}{2}N^{3}\vec{e}\,^{3}}\right)$.

As the only forces acting on the gel are applied on its boundaries,
the equilibrium of the continuous medium reads classically
\begin{equation}
{\rm div}\,\sigma =0\,.
\end{equation}
Using the virtual power formulation of that equation and the change of
variables
\begin{equation*}
(\lambda^1,\lambda^2)\leftrightarrow\vec{x}=\vec{\varphi}(\lambda^1,\lambda^2)\,,
\end{equation*}
it can be proven that the equilibrium equation reads
\begin{equation}
  \sum_{i=1}^2 \partial _{i}^{\lambda }\vec{S}^{i}=0\,.
\end{equation}
\paragraph{Coupling between growth and mechanics}
To study the growth of such an homogenized network, one can stay
within the picture of a mechanical equilibrium of the actin gel on the
time scale of the growth process.
%
%
From the homogenized elastic equations one can derive the elastic
contribution $\Delta\mu_e$ to the chemical potential $\Delta\mu$ for
the addition or subtraction of nodes at the gel interfaces starting
from the free elastic energy $F_e$ in the network, i.e.\
\begin{equation}
F_e=\frac{1}{\eta^2}\int_\Omega d\lambda^1d\lambda^2\,\sum_{b=1,2,3} f^b\,,
\end{equation}
where $f^b$ is the elastic energy associated with the extension or
contraction of each of the filaments. The elastic chemical potential
is then given by the variation of the elastic energy with respect to
the size and shape of the network $\Delta\mu_e=\delta F_e/\delta
\Omega$ by respecting the boundary conditions.

We assume now that the chemical potential contains also a contribution
from the chemical process of polymerization $\Delta\mu_c$, where
$\Delta\mu_c<0$ at the internal interface and
$\Delta\mu_c>0$ at the external interface.
This assumption accounts for the polar treadmilling behavior of the
actin polymerization, i.e.\ polymerization occurs at the internal
interface and depolymerization at the external interface.
Furthermore, the chemical potential contains a contribution from
interfacial energy, i.e.\ $\Delta\mu_s=-\gamma\kappa$, with $\gamma$
being the surface tension coefficient and $\kappa$ being the curvature
of the interface.
This leads to the following expression for the normal velocities of
the two free interfaces in the Lagrangian coordinates
$(\lambda^1,\lambda^2)$
\begin{equation}
v_n=-\eta
M^i\left(\Delta\mu^i_e+\Delta\mu^i_c+\Delta\mu^i_s\right)=-\eta
M^i\Delta\mu^i\,,
\end{equation}
with $i=0$ for the internal and $i=1$ for the external interface.
\paragraph{Homogeneous gel growth and linear stability analysis}
First one may consider the symmetric problem, i.e.\ the growth of a
gel with homogeneous thickness $\alpha=\eta N_n$ which has an
axisymmetric solution $\vec \varphi=\varphi_r(\lambda_1)\vec
e_r(\lambda_2)$.
The equilibrium equation in this case then reduces to
\begin{equation}
0=2\partial_1^\lambda\left(\tilde
N^1\partial_1^\lambda\varphi_r\right)-\left(\frac{1}{2}\tilde N^1+\tilde
N^2\right)\varphi_r\,,\label{symeq}
\end{equation}
where we have introduced
\begin{equation}
\tilde N^b=\eta\frac {N^b}{l^b}\,.
\end{equation}
In this situation the filaments with $b=2$ are oriented in a
tangential direction.  Eq.\ (\ref{symeq}) can be solved numerically
using continuation methods \cite{AUTO97}.

\begin{figure}
\begin{center}
\includegraphics[width=0.7\hsize]{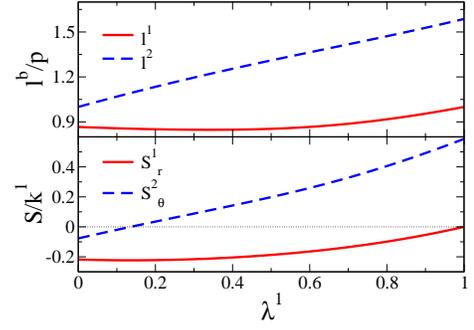}
\end{center}
\caption{\label{run3} Filament lengths (top), radial and tangential
tensions (bottom) depending on the positions in the network in
mechanical equilibrium. Parameters are $k^2/k^1=1$ and
$l^1_m=l^2_m=p$.}
\end{figure}

Fig.\ \ref{run3} shows a solution of Eq.\ (\ref{symeq}) for a given
network thickness $\alpha$.
Naturally, $l^2$ is extending as one moves away from the bead, whereas
$l^1$ is first shortening and then extending to reach its equilibrium
length at the outer gel surface.  Consequently, the gel is under
radial compression and under tangential extension far away from the
bead surface. However, for regions close to the bead surface the gel
is under tangential compression.

\begin{figure}
\begin{center}
\includegraphics[width=0.7\hsize]{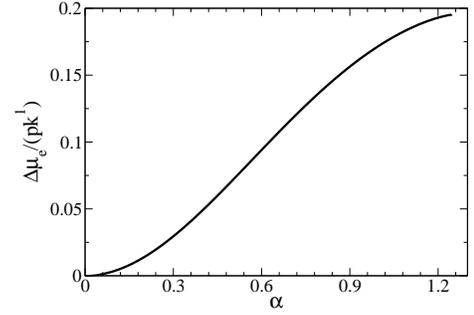}
\end{center}
\caption{\label{dmun}Dependence of the elastic part of the chemical
potential $\Delta\mu_e$ on the number of radial filament layers
$\alpha$. Remaining parameters are $l^1_m=l^2_m=p$ and $k^2/k^1=1$.}
\end{figure}

Fig.\ \ref{dmun} shows the dependence of the chemical potential
depending on the number of radial filament layers $\alpha$.
Assuming that new filaments are inserted in the same stressed state as
the already present material at the two interfaces, the elastic
chemical potential is identical at the two interfaces for a
homogeneous gel ($\Delta\mu^0_e=\Delta\mu^1_e=\Delta\mu_e$). With
increasing network size, i.e.\ increasing $\alpha$, $\Delta\mu_e$
increases in a strongly nonlinear fashion.
Note, that for higher values of $\alpha$ the homogenization approach
breaks down and the gel is starting to ``fold back''. Beyond this
point, at $\alpha \approx 1.25$ in Fig.\ \ref{dmun}, no physical
meaningful solutions exist.

We consider now a filament network which is allowed to grow
symmetrically with identical mobilities ($M^0=M^1=M$) at the two
interfaces following the two dynamic equations
\begin{eqnarray}
\partial_t\alpha^0 & = & \eta M\left(\Delta\mu^0_c+\Delta\mu_e\right)=-v_p+v_e\\
\partial_t\alpha^1 & = & -\eta M\left(\Delta\mu^1_c+\Delta\mu_e\right)=-v_d-v_e
\end{eqnarray}
whereby the positions of the internal and external interface are
denoted by $\alpha^0$ and $\alpha^1$, and where we have introduced the
polymerization speed $v_p=-\eta M\Delta\mu^0_c>0$, the
depolymerization speed $v_d=\eta M\Delta\mu^1_c$ and an ``elastic
speed'' $v_e=\eta M\Delta\mu_e$.
The steady state for the gel thickness is given by
$\partial_t\alpha=\partial_t\alpha^1-\partial_t\alpha^0=0$.
Defining now a mean velocity $\bar v=(v_p+v_d)/2$ and a velocity
difference $\Delta v=(v_p-v_d)/2$, one obtains that in the steady
state $\Delta v=v_e$ and $\partial_t\alpha^0=\partial_t\alpha^1=-\bar
v$.
This means that, although the gel thickness does not change, both
interfaces are moving with the same velocity $-\bar v$ and therefore,
$\bar v$ has the physical meaning of the treadmilling speed.

\begin{figure}
\begin{center}
\includegraphics[width=0.7\hsize]{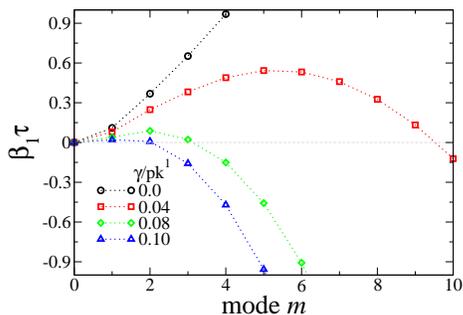}
\end{center}
\caption{\label{evals} Dispersion relation with Dirichlet boundary
conditions at the internal interface. Shown is only the larger of two
eigenvalues $\beta_1$ (the second one being always negative) depending on the
wavenumber $m$ for various values of the surface tension as indicated
in the legend. Parameters are $\alpha=1$, $k^2/k^1=1$,
$l_m^1=l_m^2=p$. The time scale is $\tau=r_0/(Mk^1p^2)$.}
\end{figure}

If we now transform the dynamical equations into the comoving frame
moving with velocity $-\bar v$ in the direction of $\lambda^1$ we can
study the linear stability of the network thickness $\alpha$ with
respect to perturbations of the type $\cos{(m\lambda^2)}$ at the
internal and external interface, $\epsilon^0(\lambda^2)$ and
$\epsilon^1(\lambda^2)$, respectively,
Fig.\ \ref{evals} shows the dispersion relation for the largest growth
rate depending on the wavenumber $m$.
One of two eigenvalues is always negative (i.e.\ stable), whereas the
other one can be positive (unstable) depending on surface tension.  We
did not find a threshold value for the filament layer number, beyond
which the gel becomes stable towards small perturbations independent
of the surface tension. However, a higher surface tension can suppress
instabilities for thin gels.
Recent experiments have shown the occurrence of higher modes than one,
i.e.\ the formation of up to three actin comets around one bead
\cite{DSR08}, depending on the experimental conditions.
The actual value for the surface tension of an actin network against
water should be rather small, since actin is a soluble
protein.
Furthermore, typically small beads with higher curvature break the
symmetry faster, than larger beads \cite{GPP05}, which is in agreement
with our model, where the time scale of symmetry-breaking increases
linearly with the radius of the bead $r_0$.
 Note also that, assuming
a constant polymerization potential, but changing the radius $r_0$ by
keeping all other parameters constant, leads to a linear relation
between the gel thickness, i.e.\ $\varphi_r(\alpha)-\varphi_r(0)$, and
$r_0$ in steady state.
These two results have also been obtained in simpler models based on
scaling arguments \cite{NGF00,PJL08,GPP05} and hold for the case that
monomer diffusion is fast enough to avoid depletion of monomers due to
polymerization at the internal bead gel interface.
Another interesting point is the type of instability one might
observe, i.e.\ an undulating vs. peristaltic instability.
For small modes $m\le 4$ one finds an undulating instability,
i.e.\ the perturbations at the external and internal interface are in
phase, whereas for higher order modes on should observe a peristaltic
instability, where the two perturbations have the opposite phase (data
not shown).
\section{Conclusions and perspectives}
In this chapter we have tried to summarize the complex properties and
out-of-equilibrium phenomena of actin gels linked by the Arp2/3
complex, which are at the origin of the motility of animal cells, as
well as of intracellular organelles and pathogens.
Primarily, we have focused on two subjects: the complex actin
polymerization dynamics under load at the polymer brush, and the
symmetry-breaking of actin gels grown from the surface of small
objects.
While we have treated both subjects separately, it is obvious that a
full understanding of the system will have to include both approaches:
the macroscopic stress distribution in the actin gel couples to the
polymerization kinetics in the polymer brush, which in turn changes
the deformation state of the gel and the macroscopic stress
distribution.
We have shown in the previous paragraphs that homogenization models
are at the moment the most appropriate models
to capture the complex microscopic structures of biological
materials on the one hand and take advantage of a continuous
framework on the other hand.
We believe, that the future in the modeling of growing actin gels in
complex geometries, e.g.\ the advancing cell edge, lies in the
coupling of these homogenization models to a complex dynamics in the
polymer brush, as proposed e.g.\ in Refs.\ \cite{EGF08,GFF08}.

An important question that remains to be elucidated is, how motion can
be generated once the gel layer has become asymmetric. So far we have
either limited our considerations to the case of a symmetry-breaking
around objects, and neglected the generation of motion, or, as in most
microscopic models, we have only considered a stationary actin comet,
which pushes an obstacle by polymerization.
Both concepts are circumventing, by more or less hand waving
arguments, one critical question. What is the origin of motion in the
absence of external forces, provided that the actin comet and the
object are only surrounded by a newtonian viscous fluid (recall that
we are in a regime with $Re\ll 1$) and not attached to some support.
Recently, Prost et al.\ \cite{PJL08} has put forward a simple argument
based on largely disparate friction coefficients for the obstacle and
the actin tail and the property of treadmilling. In the following we
will outline this argument.

Suppose that an obstacle and its associated actin comet move with
velocities ${\bf v}_o$ and ${\bf v}_c$, respectively, in the
laboratory frame. In the viscous regime the force balance reads then
\begin{equation}
0=\xi_o{\bf v}_o+\xi_c{\bf v}_c\,,\label{foba}
\end{equation}
where $\xi_o$ and $\xi_c$ denote the friction coefficients of the
object and the comet with the surrounding fluid, respectively.
The difference in the two velocities ${\bf v}_t={\bf v}_o-{\bf v}_c$
is the treadmilling speed. Substituting ${\bf v}_c$ in Eq.\ (\ref{foba})
and after some rearrangement one finds for the object velocity
\begin{equation}
{\bf v}_o={\bf v}_t\frac{\xi_c}{\xi_o+\xi_c}\,.
\end{equation}
This means, that although the dissipation due to fluid friction is
very small, it plays nevertheless a decisive role, since it is the
ratio of friction coefficients which determines the object
velocity. In the limit of $\xi_c\gg\xi_o$ this velocity approaches the
treadmilling speed.
Given the fact, that the actin comet is much larger than the object,
e.g.\ bead, droplet or vesicle, which causes a much larger friction,
the experimentally observed obstacle velocities are indeed close to
the treadmilling speed.

Another biological aspect, which might be of importance when
considering more complex cellular systems is the fact that the actin
polymerization system is not constitutively active as in {\it in
vitro} assays but is regulated by signaling cascades, which constitute
in itself a nonlinear dynamical system. Typically these signaling
cascades are modeled by reaction-diffusion systems which lead to
pattern formation \cite{MJD06}, e.g.\ polarization of the cell into
leading and advancing edge.
It remains to be shown how these two important mechanisms, elastic
instabilities and instabilities due to reaction-diffusion processes,
integrate to produce cellular motion.\\[2ex]

\noindent{\bf Acknowledgements}: C.M. and K.J. acknowledge financial
support from the CNES and the Alexander von Humboldt Foundation, and
with D.C., M.I., P.P., and A.R. we acknowledge a financial support
from ANR MOSICOB (MOd\'elisation et SImulation de fluides COmplexes
Biomim\'etiques).
%
%
%

\begin{thebibliography}{10}
\expandafter\ifx\csname bibinfo\endcsname\relax\def\bibinfo#1#2{#2}\fi

\bibitem{Abe80}
\bibinfo{author}{M.~Abercrombie} (\bibinfo{year}{1980}) \bibinfo{title}{The
  crawling movement of metazoan cells}. {\it \bibinfo{journal}{Proc. R. Soc.
  London B}\/} {\bf \bibinfo{volume}{207}}:\bibinfo{pages}{129--147},
  \bibinfo{note}{and references therein}.

\bibitem{AlO04}
\bibinfo{author}{J.B.~Alberts} and \bibinfo{author}{G.M.~Odell}
  (\bibinfo{year}{2004}) \bibinfo{title}{In silico reconstitution of {\it
  Listeria} propulsion exhibits nano-saltation}. {\it \bibinfo{journal}{PloS
  Biol.}\/} {\bf \bibinfo{volume}{2}}:\bibinfo{pages}{e412}.

\bibitem{AnC00} \bibinfo{author}{K.I.~Anderson} and
\bibinfo{author}{R.~Cross} (\bibinfo{year}{2000})
\bibinfo{title}{Contact dynamics during keratocyte motility}. {\it
\bibinfo{journal}{Curr. Biol.}\/} {\bf
\bibinfo{volume}{10}}:\bibinfo{pages}{253--260}.

\bibitem{BPS05} \bibinfo{author}{A.~Bernheim-Groswasser},
\bibinfo{author}{J.~Prost} and \bibinfo{author}{C.~Sykes}
(\bibinfo{year}{2005}) \bibinfo{title}{Mechanism of actin-based
motility: A dynamic state diagram}. {\it
\bibinfo{journal}{Biophys. J.}\/} {\bf
\bibinfo{volume}{89}}:\bibinfo{pages}{1411--1419}.

\bibitem{BWG02}
\bibinfo{author}{A.~Bernheim-Groswasser}, \bibinfo{author}{S.~Wiesner},
  \bibinfo{author}{R.M.~Golsteyn}, \bibinfo{author}{M.-F.~Carlier} and
  \bibinfo{author}{C.~Sykes} (\bibinfo{year}{2002}) \bibinfo{title}{The
  dynamics of actin-based motility depend on surface parameters}. {\it
  \bibinfo{journal}{Nature}\/} {\bf \bibinfo{volume}{417}}:\bibinfo{pages}{308--311}.

\bibitem{BCJ04}
\bibinfo{author}{H.~Boukellal}, \bibinfo{author}{O.~Camp\'as},
  \bibinfo{author}{J.-F.~Joanny}, \bibinfo{author}{J.~Prost} and
  \bibinfo{author}{C.~Sykes} (\bibinfo{year}{2004}) \bibinfo{title}{Soft {\it
  Listeria}: Actin-based propulsion of liquid drops}. {\it
  \bibinfo{journal}{Phys. Rev. E}\/} {\bf \bibinfo{volume}{69}}:\bibinfo{pages}{061906}.

\bibitem{Bray}
\bibinfo{author}{D.~Bray} (\bibinfo{year}{1992}) {\it \bibinfo{title}{Cell
  movements}\/},  \bibinfo{address}{New York \& London}, \bibinfo{publisher}{Garland Publishing, Inc.}.

\bibitem{BuM07}
\bibinfo{author}{N.J.~Burroughs} and \bibinfo{author}{D.~Marenduzzo}
  (\bibinfo{year}{2007}) \bibinfo{title}{Nonequilibrium-driven motion in actin
  networks: {C}omet tails and moving beads}. {\it \bibinfo{journal}{Phys. Rev.
  Lett.}\/} {\bf \bibinfo{volume}{98}}:\bibinfo{pages}{238302}.

\bibitem{CaC01} 
\bibinfo{author}{D.~Caillerie} and
\bibinfo{author}{B.~Cambou} (\bibinfo{year}{2001})
\bibinfo{chapter}{Les techniques de changement d'\'echelles dans les
mat\'eriaux granulaires}, in {\it \bibinfo{title}{Microm\'ecanique des milieux
  granulaires}\/}, \bibinfo{address}{Paris},
  \bibinfo{publisher}{Herm\`es Sciences}.

\bibitem{CJM09}
\bibinfo{author}{D.~Caillerie}, \bibinfo{author}{K.~John},
  \bibinfo{author}{N.~Meunier}, \bibinfo{author}{C.~Misbah},
  \bibinfo{author}{P.~Peyla} and \bibinfo{author}{A.~Raoult}
  (\bibinfo{year}{2009}) \bibinfo{title}{A model for actin driven motility
  through discrete homogenization}. {\it \bibinfo{journal}{manuscript in
  preparation}\/} .

\bibitem{CFO99}
\bibinfo{author}{L.A.~Cameron}, \bibinfo{author}{M.J.~Footer},
  \bibinfo{author}{A.~Van~Oudenaarden} and \bibinfo{author}{J.A.~Theriot}
  (\bibinfo{year}{1999}) \bibinfo{title}{Motility of ActA protein-coated
  microspheres driven by actin polymerization}. {\it \bibinfo{journal}{Proc.
  Natl Acad. Sci. USA}\/} {\bf \bibinfo{volume}{96}}:\bibinfo{pages}{4908--4913}.

\bibitem{Car01}
\bibinfo{author}{A.E.~Carlsson} (\bibinfo{year}{2001}) \bibinfo{title}{Growth
  of branched actin networks against obstacles}. {\it
  \bibinfo{journal}{Biophys. J.}\/} {\bf \bibinfo{volume}{81}}:\bibinfo{pages}{1907--1923}.

\bibitem{CCG95}
\bibinfo{author}{S.~Cudmore}, \bibinfo{author}{P.~Cossart},
  \bibinfo{author}{G.~Griffiths} and \bibinfo{author}{M.~Way}
  (\bibinfo{year}{1995}) \bibinfo{title}{Actin-based motility of {\it Vaccinia}
  virus}. {\it \bibinfo{journal}{Nature}\/} {\bf \bibinfo{volume}{378}}:\bibinfo{pages}{636--638}.

\bibitem{DSR08}
\bibinfo{author}{V.~Delatour}, \bibinfo{author}{S.~Shekhar},
  \bibinfo{author}{A.-C.~Reymann}, \bibinfo{author}{D.~Didry},
  \bibinfo{author}{K.~H\^o Di\^ep~L\^e}, \bibinfo{author}{G.~Romet-Lemonne},
  \bibinfo{author}{E.~Helfer} and \bibinfo{author}{M.-F.~Carlier}
  (\bibinfo{year}{2008}) \bibinfo{title}{Actin-based propulsion of
  functionalized hard versus fluid spherical objects}. {\it
  \bibinfo{journal}{New J. Phys.}\/} {\bf \bibinfo{volume}{10}}:\bibinfo{pages}{025001}.

\bibitem{AUTO97}
\bibinfo{author}{{E.J.~Doedel}}, \bibinfo{author}{{A.R.~Champneys}},
  \bibinfo{author}{{T.F.~Fairgrieve}}, \bibinfo{author}{{Y.A.~Kuznetsov}},
  \bibinfo{author}{B.~Sandstede} and \bibinfo{author}{{X.J.~Wang}}
  (\bibinfo{year}{1997}) {\it \bibinfo{title}{AUTO97: Continuation and
  bifurcation software for ordinary differential equations}\/}, \bibinfo{address}{Montreal},
  \bibinfo{publisher}{Concordia University}.

\bibitem{EGF08}
\bibinfo{author}{M.~Enculescu}, \bibinfo{author}{A.~Gholami} and
  \bibinfo{author}{M.~Falcke} (\bibinfo{year}{2008}) \bibinfo{title}{Dynamic
  regimes and bifurcations in a model of actin-based motility}. {\it
  \bibinfo{journal}{Phys. Rev. E}\/} {\bf \bibinfo{volume}{78}}:\bibinfo{pages}{031915}.

\bibitem{FlT04}
\bibinfo{author}{D.A.~Fletcher} and \bibinfo{author}{J.A.~Theriot}
  (\bibinfo{year}{2004}) \bibinfo{title}{An introduction to cell motility for
  the physical scientist}. {\it \bibinfo{journal}{Phys. Biol.}\/} {\bf
  \bibinfo{volume}{1}}:\bibinfo{pages}{T1--T10}.

\bibitem{FuI97}
\bibinfo{author}{Y.~Fukui} and \bibinfo{author}{S.~Inou\'e}
  (\bibinfo{year}{1997}) \bibinfo{title}{Amoeboid movement anchored by eupodia,
  new actin-rich knobby feet in {\it Dictyostelium}}. {\it
  \bibinfo{journal}{Cell Motil. Cytoskel.}\/} {\bf \bibinfo{volume}{36}}:\bibinfo{pages}{339--354}.

\bibitem{GCR00}
\bibinfo{author}{F.~Gerbal}, \bibinfo{author}{P.~Chaikin},
  \bibinfo{author}{Y.~Rabin} and \bibinfo{author}{J.~Prost}
  (\bibinfo{year}{2000}) \bibinfo{title}{An elastic analysis of {\it Listeria
  monocytogenes} propulsion}. {\it \bibinfo{journal}{Biophys. J.}\/} {\bf
  \bibinfo{volume}{79}}:\bibinfo{pages}{2259--2275}.

\bibitem{GLO00}
\bibinfo{author}{F.~Gerbal}, \bibinfo{author}{V.~Laurent},
  \bibinfo{author}{A.~Ott}, \bibinfo{author}{M.-F.~Carlier},
  \bibinfo{author}{P.~Chaikin} and \bibinfo{author}{J.~Prost}
  (\bibinfo{year}{2000}) \bibinfo{title}{Measurement of the elasticity of the
  actin tail of {\it Listeria monocytogenes}}. {\it \bibinfo{journal}{Eur.
  Biophys. J.}\/} {\bf \bibinfo{volume}{29}}:\bibinfo{pages}{134--140}.

\bibitem{GFF08}
\bibinfo{author}{A.~Gholami}, \bibinfo{author}{M.~Falcke} and
  \bibinfo{author}{E.~Frey} (\bibinfo{year}{2008}) \bibinfo{title}{Velocity
  oscillations in actin-based motility}. {\it \bibinfo{journal}{New J.
  Phys.}\/} {\bf \bibinfo{volume}{10}}:\bibinfo{pages}{033022}.

\bibitem{GWF06}
\bibinfo{author}{A.~Gholami}, \bibinfo{author}{J.~Wilhelm} and
  \bibinfo{author}{E.~Frey} (\bibinfo{year}{2006}) \bibinfo{title}{Entropic
  forces generated by grafted semiflexible polymers}. {\it
  \bibinfo{journal}{Phys. Rev. E}\/} {\bf
  \bibinfo{volume}{74}}:\bibinfo{pages}{041803}.

\bibitem{GFT03}
\bibinfo{author}{P.A.~Giardini}, \bibinfo{author}{D.A.~Fletcher} and
  \bibinfo{author}{J.A.~Theriot} (\bibinfo{year}{2003})
  \bibinfo{title}{Compression forces generated by actin comet tails on lipid
  vesicles}. {\it \bibinfo{journal}{Proc. Nat. Acad. Sci. USA}\/} {\bf
  \bibinfo{volume}{100}}:\bibinfo{pages}{6493--6498}.

\bibitem{GoT95} \bibinfo{author}{M.B.~Goldberg} and
\bibinfo{author}{J.A.~Theriot} (\bibinfo{year}{1995})
\bibinfo{title}{{\it Shigella flexneri} surface protein {IcsA} is
sufficient to direct actin-based motility}. {\it
\bibinfo{journal}{Proc. Natl Acad. Sci. USA}\/} {\bf
\bibinfo{volume}{92}}:\bibinfo{pages}{6572--6576}.

\bibitem{HPG90}
\bibinfo{author}{K.C.~Holmes}, \bibinfo{author}{D.~Popp},
  \bibinfo{author}{W.~Gebhard} and \bibinfo{author}{W.~Kabsch}
  (\bibinfo{year}{1990}) \bibinfo{title}{Atomic model of the actin filament}.
  {\it \bibinfo{journal}{Nature}\/} {\bf \bibinfo{volume}{347}}:\bibinfo{pages}{44--49}.

\bibitem{Howard}
\bibinfo{author}{J.~Howard} (\bibinfo{year}{2001}) {\it
  \bibinfo{title}{Mechanics of motor proteins and the cytoskeleton}\/},
  \bibinfo{publisher}{Sinauer Associates, Inc.}

\bibitem{IRH76} \bibinfo{author}{G.~Isenberg},
\bibinfo{author}{P.C.~Rathke}, \bibinfo{author}{N.~H\"ulsmann},
\bibinfo{author}{W.W.~Franke} and
\bibinfo{author}{K.E.~Wohlfahrt-Bottermann} (\bibinfo{year}{1976})
\bibinfo{title}{Cytoplasmic actomyosin fibrils in tissue culture
cells.  {D}irect proof of contractility by visualization of
{ATP}-induced contraction in fibrils isolated by laser microbeam
dissection.} {\it \bibinfo{journal}{Cell Tiss. Res.}\/} {\bf
\bibinfo{volume}{166}}:\bibinfo{pages}{427--443}.

\bibitem{JPK08}
\bibinfo{author}{K.~John}, \bibinfo{author}{P.~Peyla},
  \bibinfo{author}{K.~Kassner}, \bibinfo{author}{J.~Prost} and
  \bibinfo{author}{C.~Misbah} (\bibinfo{year}{2008}) \bibinfo{title}{A
  nonlinear study of symmetry-breaking in actin gels - {I}mplications for
  cellular motility}. {\it \bibinfo{journal}{Phys. Rev. Lett.}\/} {\bf
  \bibinfo{volume}{100}}:\bibinfo{pages}{068101}.

\bibitem{KTD06}
\bibinfo{author}{M.~Kaksonen}, \bibinfo{author}{C.P.~Toret} and
  \bibinfo{author}{D.G.~Drubin} (\bibinfo{year}{2006})
  \bibinfo{title}{Harnessing actin dynamics for chlathrin-mediated
  endocytosis}. {\it \bibinfo{journal}{Nat. Rev. Mol. Cell. Biol.}\/} {\bf
  \bibinfo{volume}{7}}:\bibinfo{pages}{404--414}.

\bibitem{KrS01}
\bibinfo{author}{M.~Krupa} and \bibinfo{author}{P.~Szmolyan}
  (\bibinfo{year}{2001}) \bibinfo{title}{Relaxation oscillation and canard
  explosion}. {\it \bibinfo{journal}{J. Differ. Equations}\/} {\bf
  \bibinfo{volume}{174}}:\bibinfo{pages}{312 -- 368}.

\bibitem{LGG97}
\bibinfo{author}{I.~Lasa}, \bibinfo{author}{E.~Gouin},
  \bibinfo{author}{M.~Goethals}, \bibinfo{author}{K.~Vancompernolle},
  \bibinfo{author}{V.~David}, \bibinfo{author}{J.~Vandekerckhove} and
  \bibinfo{author}{P.~Cossart} (\bibinfo{year}{1997})
  \bibinfo{title}{Identification of two regions in the N-terminal domain of
  {ActA} involved in the actin comet tail formation by {\it Listeria
  monocytogenes}}. {\it \bibinfo{journal}{EMBO J.}\/} {\bf
  \bibinfo{volume}{16}}:\bibinfo{pages}{1531--1540}.

\bibitem{LLK05} \bibinfo{author}{A.~Lee}, \bibinfo{author}{H.Y.~Lee}
and \bibinfo{author}{M.~Kardar} (\bibinfo{year}{2005})
\bibinfo{title}{Symmetry-breaking motility}. {\it
\bibinfo{journal}{Phys.  Rev. Lett.}\/} {\bf
\bibinfo{volume}{95}}:\bibinfo{pages}{138101}.

\bibitem{LBP99}
\bibinfo{author}{T.P.~Loisel}, \bibinfo{author}{R.~Boujemaa},
  \bibinfo{author}{D.~Pantaloni} and \bibinfo{author}{M.-F.~Carlier}
  (\bibinfo{year}{1999}) \bibinfo{title}{Reconstitution of actin-based motility
  of {\it Listeria} and {\it Shigella} using pure proteins}. {\it
  \bibinfo{journal}{Nature}\/} {\bf \bibinfo{volume}{401}}:\bibinfo{pages}{613--616}.

\bibitem{Lov44}
\bibinfo{author}{A.E.H.~Love} (\bibinfo{year}{1944}) {\it \bibinfo{title}{A
  treatise of the mathematical theory of elasticity}\/}, \bibinfo{address}{New York},
  \bibinfo{publisher}{Dover Publications}.

\bibitem{MCJ98}
\bibinfo{author}{L.~Ma}, \bibinfo{author}{L.C.~Cantley},
  \bibinfo{author}{P.A.~Janmey} and \bibinfo{author}{M.W.~Kirschner}
  (\bibinfo{year}{1998}) \bibinfo{title}{Corequirement of specific
  phosphoinositides and small {GTP}-binding protein {C}dc42 in inducing actin
  assembly in {\it Xenopus} egg extracts}. {\it \bibinfo{journal}{J. Cell
  Biol.}\/} {\bf \bibinfo{volume}{140}}:\bibinfo{pages}{1125--1136}.

\bibitem{MPC04} \bibinfo{author}{Y.~Marcy},
\bibinfo{author}{J.~Prost}, \bibinfo{author}{M.-F.~Carlier} and
\bibinfo{author}{C.~Sykes} (\bibinfo{year}{2004})
\bibinfo{title}{Forces generated during actin-based propulsion: A
direct measurement by micromanipulation}. {\it
\bibinfo{journal}{Proc. Natl Acad. Sci. USA}\/} {\bf
\bibinfo{volume}{101}}:\bibinfo{pages}{5992--5997}.

\bibitem{MJD06}
\bibinfo{author}{A.F.M.~Mar\'ee}, \bibinfo{author}{A.~Jilkine},
  \bibinfo{author}{A.~Dawes}, \bibinfo{author}{V.A.~Grieneisen} and
  \bibinfo{author}{L.~Edelstein-Keshet} (\bibinfo{year}{2006})
  \bibinfo{title}{Polarization and movement of keratocytes: A multiscale
  modelling approach}. {\it \bibinfo{journal}{Bull. Math. Biol.}\/} {\bf
  \bibinfo{volume}{68}}:\bibinfo{pages}{1169--1211}.

\bibitem{MBG07}
\bibinfo{author}{A.~Michelot}, \bibinfo{author}{J.~Berro},
  \bibinfo{author}{C.~Gu\'erin}, \bibinfo{author}{R.~Boujemaa-Paterski},
  \bibinfo{author}{C.~J. Staiger}, \bibinfo{author}{J.-L.~Martiel} and
  \bibinfo{author}{L.~Blanchoin} (\bibinfo{year}{2007})
  \bibinfo{title}{Actin-filament stochastic dynamics mediated by
  {ADF/Cofilin}}. {\it \bibinfo{journal}{Curr. Biol.}\/} {\bf
  \bibinfo{volume}{17}}:\bibinfo{pages}{825--833}.

\bibitem{MoO96}
\bibinfo{author}{A.~Mogilner} and \bibinfo{author}{G.~Oster}
  (\bibinfo{year}{1996}) \bibinfo{title}{Cell motility driven by actin
  polymerization}. {\it \bibinfo{journal}{Biophys. J.}\/} {\bf
  \bibinfo{volume}{71}}:\bibinfo{pages}{3030--3045}.

\bibitem{MoO03} \bibinfo{author}{A.~Mogilner} and
\bibinfo{author}{G.~Oster} (\bibinfo{year}{2003})
\bibinfo{title}{Force generation by actin polymerization II: The
elastic ratchet and tethered filaments}. {\it
\bibinfo{journal}{Biophys. J.}\/} {\bf
\bibinfo{volume}{84}}:\bibinfo{pages}{1591--1605}.

\bibitem{NGF00}
\bibinfo{author}{V.~Noireaux}, \bibinfo{author}{R.M.~Golsteyn},
  \bibinfo{author}{E.~Friederich}, \bibinfo{author}{J.~Prost},
  \bibinfo{author}{C.~Antony}, \bibinfo{author}{D.~Louvard} and
  \bibinfo{author}{C.~Sykes} (\bibinfo{year}{2000}) \bibinfo{title}{Growing an
  actin gel on spherical surfaces}. {\it \bibinfo{journal}{Biophys. J.}\/} {\bf
  \bibinfo{volume}{78}}:\bibinfo{pages}{1643--1654}.

\bibitem{OMS93}
\bibinfo{author}{A.~Ott}, \bibinfo{author}{M.~Magnasco},
  \bibinfo{author}{A.~Simon} and \bibinfo{author}{A.~Libchaber}
  (\bibinfo{year}{1993}) \bibinfo{title}{Measurement of the persistence length
  of polymerized actin using fluorescence microscopy}. {\it
  \bibinfo{journal}{Phys. Rev. E}\/} {\bf \bibinfo{volume}{48}}:\bibinfo{pages}{R1642--R1645}.

\bibitem{PCT05}
\bibinfo{author}{S.H.~Parekh}, \bibinfo{author}{O.~Chaudhuri},
  \bibinfo{author}{J.~A. Theriot} and \bibinfo{author}{D.~A. Fletcher}
  (\bibinfo{year}{2005}) \bibinfo{title}{Loading history determines the
  velocity of actin-network growth}. {\it \bibinfo{journal}{Nature Cell
  Biol.}\/} {\bf \bibinfo{volume}{7}}:\bibinfo{pages}{1219--1223}.

\bibitem{POO93}
\bibinfo{author}{C.S.~Peskin}, \bibinfo{author}{G.M.~Odell} and
  \bibinfo{author}{G.F.~Oster} (\bibinfo{year}{1993}) \bibinfo{title}{Cellular
  motions and thermal fluctuations: The brownian ratchet}. {\it
  \bibinfo{journal}{Biophys. J.}\/} {\bf \bibinfo{volume}{65}}:\bibinfo{pages}{316--324}.

\bibitem{PLP04} \bibinfo{author}{J.~Plastino},
\bibinfo{author}{I.~Lelidis}, \bibinfo{author}{J.~Prost} and
\bibinfo{author}{C.~Sykes} (\bibinfo{year}{2004}) \bibinfo{title}{The
effect of diffusion, depolymerization and nucleation promoting factors
on actin gel growth}. {\it \bibinfo{journal}{Eur. Biophys. J.}\/} {\bf
\bibinfo{volume}{33}}:\bibinfo{pages}{310--320}.

\bibitem{PBM00}
\bibinfo{author}{T.D.~Pollard}, \bibinfo{author}{L.~Blanchoin} and
  \bibinfo{author}{R.D.~Mullins} (\bibinfo{year}{2000})
  \bibinfo{title}{Molecular mechanisms controlling actin filament dynamics in
  nonmuscle cells}. {\it \bibinfo{journal}{Ann. Rev. Biophys. Biomol.
  Struct.}\/} {\bf \bibinfo{volume}{29}}:\bibinfo{pages}{545--576}.

\bibitem{PoB03}
\bibinfo{author}{T.D.~Pollard} and \bibinfo{author}{G.G.~Borisy}
  (\bibinfo{year}{2003}) \bibinfo{title}{Cellular motility driven by assembly
  and disassembly of actin filaments}. {\it \bibinfo{journal}{Cell}\/} {\bf
  \bibinfo{volume}{112}}:\bibinfo{pages}{453--465}.


\bibitem{PJL08}
\bibinfo{author}{J.~Prost}, \bibinfo{author}{J.-F.~Joanny},
  \bibinfo{author}{P.~Lenz} and \bibinfo{author}{C.~Sykes}
  (\bibinfo{year}{2008}) \bibinfo{chapter}{The physics of {\it Listeria} propulsion}, in 
 {\it \bibinfo{title}{Cell Motility}\/}, \bibinfo{pages}{1--30}, \bibinfo{address}{New
  York}, \bibinfo{publisher}{Springer}.

\bibitem{RaT04} \bibinfo{author}{S.M.~Rafelski} and
\bibinfo{author}{J.A.~Theriot} (\bibinfo{year}{2004})
\bibinfo{title}{Crawling toward a unified model of cell motility:
Spatial and temporal regulation of actin dynamics}. {\it
\bibinfo{journal}{Annu. Rev. Biochem.}\/} {\bf
\bibinfo{volume}{73}}:\bibinfo{pages}{209--239}.

\bibitem{SPJ04}
\bibinfo{author}{K.~Sekimoto}, \bibinfo{author}{J.~Prost},
  \bibinfo{author}{F.~J\"ulicher}, \bibinfo{author}{H.~Boukellal} and
  \bibinfo{author}{A.~Bernheim-Grosswasser} (\bibinfo{year}{2004})
  \bibinfo{title}{Role of tensile stress in actin gels and a symmetry-breaking
  instability}. {\it \bibinfo{journal}{Eur. Phys. J. E}\/} {\bf
  \bibinfo{volume}{13}}:\bibinfo{pages}{247--259}.

\bibitem{Stryer}
\bibinfo{author}{L.~Stryer} (\bibinfo{year}{1995}) {\it
  \bibinfo{title}{Biochemistry}\/}, \bibinfo{address}{New York}, \bibinfo{publisher}{W.~H.~Freeman \&
  Company}.

\bibitem{ThM91}
\bibinfo{author}{J.A.~Theriot} and \bibinfo{author}{T.J.~Mitchison}
  (\bibinfo{year}{1991}) \bibinfo{title}{Actin microfilament dynamics in
  locomoting cells}. {\it \bibinfo{journal}{Nature}\/} {\bf
  \bibinfo{volume}{352}}:\bibinfo{pages}{126--131}.

\bibitem{TMT92}
\bibinfo{author}{T.A.~Theriot}, \bibinfo{author}{T.J.~Mitchison},
  \bibinfo{author}{L.G.~Tilney} and \bibinfo{author}{D.A.~Portnoy}
  (\bibinfo{year}{1992}) \bibinfo{title}{The rate of actin-based motility of
  intracellular {\it Listeria monocytogenes} equals the rate of actin
  polymerization}. {\it \bibinfo{journal}{Nature}\/} {\bf
  \bibinfo{volume}{357}}:\bibinfo{pages}{257--260}.

\bibitem{ToC98}
\bibinfo{author}{H.~Tollenaere} and \bibinfo{author}{D.~Caillerie}
  (\bibinfo{year}{1998}) \bibinfo{title}{Continuous modeling of lattice
  structures by homogenization}. {\it \bibinfo{journal}{Adv. Eng. Software}\/}
  {\bf \bibinfo{volume}{29}}:\bibinfo{pages}{699--705}.

\bibitem{TCS07}
\bibinfo{author}{L.~Trichet}, \bibinfo{author}{O.~Camp\`as},
  \bibinfo{author}{C.~Sykes} and \bibinfo{author}{J.~Plastino}
  (\bibinfo{year}{2007}) \bibinfo{title}{{VASP} governs actin dynamics by
  modulating filament anchoring}. {\it \bibinfo{journal}{Biophys. J.}\/} {\bf
  \bibinfo{volume}{92}}:\bibinfo{pages}{1081--1089}.

\bibitem{UCA03}
\bibinfo{author}{A.~Upadhyaya}, \bibinfo{author}{J.R.~Chabot},
  \bibinfo{author}{A.~Andreeva}, \bibinfo{author}{A.~Samadani} and
  \bibinfo{author}{A.~van Oudenaarden} (\bibinfo{year}{2003})
  \bibinfo{title}{Probing polymerization forces by using actin-propelled lipid
  vesicles}. {\it \bibinfo{journal}{Proc. Nat. Acad. Sci. USA}\/} {\bf
  \bibinfo{volume}{100}}:\bibinfo{pages}{4521--4526}.

\bibitem{UKG04} \bibinfo{author}{M.~Urbakh},
\bibinfo{author}{J.~Klafter}, \bibinfo{author}{D.~Gourdon} and
\bibinfo{author}{J.~Israelachvili} (\bibinfo{year}{2004})
\bibinfo{title}{The nonlinear nature of friction}.  {\it
\bibinfo{journal}{Nature}\/} {\bf
\bibinfo{volume}{430}}:\bibinfo{pages}{525--528}.

\bibitem{GPP05}
\bibinfo{author}{J.~van~der Gucht}, \bibinfo{author}{E.~Paluch},
  \bibinfo{author}{J.~Plastino} and \bibinfo{author}{C.~Sykes}
  (\bibinfo{year}{2005}) \bibinfo{title}{Stress release drives symmetry
  breaking for actin-based movement}. {\it \bibinfo{journal}{Proc. Natl Acad.
  Sci. USA}\/} {\bf \bibinfo{volume}{102}}:\bibinfo{pages}{7847--7852}.

\bibitem{OuT99}
\bibinfo{author}{A.~van Oudenaarden} and \bibinfo{author}{J.A.~Theriot}
  (\bibinfo{year}{1999}) \bibinfo{title}{Cooperative symmetry-breaking by actin
  polymerization in a model for cell motility}. {\it \bibinfo{journal}{Nature
  Cell Biol.}\/} {\bf \bibinfo{volume}{1}}:\bibinfo{pages}{493--499}.

\bibitem{VSB99} \bibinfo{author}{A.B.~Verkhovsky},
\bibinfo{author}{T.M.~Svitkina} and \bibinfo{author}{G.G.~Borisy}
(\bibinfo{year}{1999}) \bibinfo{title}{Self-polarization and
directional motility of cytoplasm}.  {\it
\bibinfo{journal}{Curr. Biol.}\/} {\bf
\bibinfo{volume}{9}}:\bibinfo{pages}{11--20}.

\bibitem{WHD03}
\bibinfo{author}{S.~Wiesner}, \bibinfo{author}{E.~Helfer},
  \bibinfo{author}{D.~Didry}, \bibinfo{author}{G.~Ducouret},
  \bibinfo{author}{F.~Lafuma}, \bibinfo{author}{M.-F.~Carlier} and
  \bibinfo{author}{D.~Pantaloni} (\bibinfo{year}{2003}) \bibinfo{title}{A
  biomimetic motility assay provides insight into the mechanism of actin-based
  motility}. {\it \bibinfo{journal}{J. Cell Biol.}\/} {\bf
  \bibinfo{volume}{160}}:\bibinfo{pages}{387--398}.

\bibitem{YTA99} \bibinfo{author}{D.~Yarar}, \bibinfo{author}{W.~To},
\bibinfo{author}{A.~Abo} and \bibinfo{author}{M.D.~Welch}
(\bibinfo{year}{1999}) \bibinfo{title}{The {Wiskott-Aldrich} syndrome
protein directs actin-based motility by stimulating actin nucleation
with the {Arp2/3} complex}. {\it \bibinfo{journal}{Curr. Biol.}\/}
{\bf \bibinfo{volume}{9}}:\bibinfo{pages}{555--558}.

\end{thebibliography}
%
%

%
\end{document}